\documentclass[final]{elsarticle}

\usepackage{lineno, hyperref, pifont, natbib, geometry, graphicx}
\renewcommand{\sectionautorefname}
{\ifnum \spacefactor > 1000 Section \else Section \fi}

\journal{}
\usepackage[USenglish]{babel}
\usepackage{amsmath}
\usepackage{amsfonts}
\usepackage{amssymb}
\usepackage{mathrsfs}
\usepackage{multirow}
\usepackage{cases}
\usepackage{subfig}
\usepackage{booktabs}
\usepackage{hyperref}
\usepackage{xcolor}
\usepackage{float}
\usepackage{algorithm} 
\usepackage{comment}
\usepackage{hhline}
\usepackage{booktabs}
\usepackage[]{draftfigure}
\usepackage{algpseudocode}

\let\today\relax
\makeatletter
\def\ps@pprintTitle{%
    \let\@oddhead\@empty
    \let\@evenhead\@empty
    \def\@oddfoot{\footnotesize\itshape
         {Preprint adapted for arXiv} \hfill\today}%
    \let\@evenfoot\@oddfoot
    }
\makeatother

\begin{document}

\begin{frontmatter}

\title{
Optimization of the initial post-buckling response
of trusses and frames by an asymptotic approach
}

\author[dtu]{Federico Ferrari \fnref{ft1}\corref{cor1}}
 \ead{feferr@dtu.dk}
\author[dtu]{Ole Sigmund \fnref{ft2}}
 \ead{olsi@dtu.dk}

\fntext[ft1]{Researcher}
\fntext[ft2]{Full Professor}

\cortext[cor1]{Corresponding author}

\address[dtu]{Department of Civil \& Mechanical Engineering, Technical University of Denmark, Koppels All\'{e} 404, 2800 Kongens Lyngby, Denmark}

\begin{abstract}
Asymptotic post-buckling theory is applied to sizing and topology optimization of trusses and frames, exploring its potential and current computational difficulties. We show that a designs' post-buckling response can be controlled by including the lowest two asymptotic coefficients, representing the initial post-buckling slope and curvature, in the optimization formulation. This also reduces the imperfection sensitivity of the optimized design. The asymptotic expansion can further be used to approximate the structural nonlinear response, and then to optimize for a given measure of the nonlinear mechanical performance such as, for example, end-compliance or complementary work. Examples of linear and nonlinear compliance minimization of trusses and frames show the effective use of the asymptotic method for including post-buckling constraints in structural optimization.
\end{abstract}
\end{frontmatter}

\section{Introduction}
 \label{Sec:Introduction}

Maximum stiffness or minimum stress design often leads to structural assemblies with slender members dominated by axial forces \cite{thompson-supple_73a}, which are prone to simultaneously reaching their critical load. This may cause a complex buckling behaviour and mode interaction, which in turn amplifies imperfection sensitivity and reduces the actual limit load \cite{ho-72a_influenceOfImperfections, ho-74a_bucklingLoadsMultipleEigenvalues}. Thus, it is important not only to accurately compute the structures' actual buckling strength, but also to assess the stability of their initial post-buckling responses, while considering all possible imperfections.

Post-buckling response is non-trivial to simulate, and in the last decades much research has been focused on developing robust and accurate analysis tools. Including post-buckling response within structural optimization poses formidable challenges, to date not yet resolved satisfactorily neither by introducing buckling constraints, nor by optimizing for nonlinear response functions.

Buckling constraints \cite{ferrari-sigmund_19a_revisitingTObuckling, ferrari-sigmund_20a} only affect the limit load of the optimized design, but neither give rigorous control of its post-buckling response nor on its imperfection sensitivity. Although some studies have shown that structures designed with buckling constraints can withstand a larger load when considering nonlinear effects \cite{ferrari-sigmund_23a_regularizationLocalizedBuckling, bluhm-etal_22a_experimentalVerificationHierarchicalLattices}, this is not a rule, especially when introducing imperfections. On the other hand, optimization of nonlinear response functions, such as end-compliance, end-stiffness, complementary work, etc. \cite{buhl-etal_00a_geometricNonlinearTO, kemmler-etal_05a_largeDeformationStability, dalklint-etal_20a_eigenfrequencyHyperelastic}, may give better control of the pre-buckling regime, but still no information on the buckling mechanism.

Control of the buckling mechanism, and of the structural stability afterwards, requires computing the initial post-buckling response. This is classically performed by potentially complicated and expensive path-following procedures, based on incremental iterative schemes \cite{book:krenk2009, book:deborst2012}. These quickly become computationally infeasible for repeated analysis tasks, as often required by large-scale topology optimization (TO). Thus, very few works have dealt with structural TO accounting for nonlinear stability, or the initial post-buckling response \cite{li-etal_22a_freeFormMultimaterialSnapping, zhang-etal_23a_finiteStrainTOstability}.

The asymptotic post-buckling theory, developed by Budiansky \cite{budiansky_74a_theoryBucklingPostBuckling} building on the  perturbation approach initiated by Koiter \cite{koiter_45a_stabilityElasticEquilibrium}, is an elegant and effective alternative for the investigation of the initial post-buckling response. In the asymptotic approach, the nonlinear response in the vicinity of a critical point is approximated by using higher order variations of the equilibrium equations. The approximation is a function of displacement fields, the load parameter, and of a set of asymptotic coefficients, the lowest of which give direct information about the nature of the bifurcation point.

After some initial criticism \cite{gallagher_75a_perturbationProceduresNonlinearFEA}, the past thirty years have seen extensive developments in the adoption of Koiters' formalism within the computational setting, resulting in a family of Asymptotic-Numerical Methods (ANM), combining the power of analytical (asymptotic) expressions, and the flexibility of numerically computed coefficients.

However, the accuracy and reliability of the ANM approximation crucially relies on avoiding spurious effects due to: (1) the use of non-objective, kinematically constrained structural models, (2) the occurrence of interpolation, and (3) extrapolation locking (see \cite{NOTE_garcea-etal_13a_postBucklingAsymptotic} for a detailed review of these phenomena). The importance of using objective structural models, not polluted by spurious deformations at large rotations, was pointed out by \cite{pignataro-etal_82a_nonlinearBeamModelPostbuckling}, and a solution to this was provided by the implicit co-rotational formulation \cite{garcea-etal_12a_implicitCoRotationalModel, garcea-etal_12b_nonFEManalysisImplicitCoRotational}. The interpolation locking phenomenon, first acknowledged by \cite{lanzo-etal_95a_asymptoticPostBucklingPlates}, affects the curvature of the post-buckling path, due to a wrong estimate of the second post-buckling coefficient. This was characterized by \cite{salerno-lanzo_97a_nonlinearFElockingFree} and then solved based on a new theoretical definition of the deformation measure, following the guidelines of \cite{pignataro-etal_82a_nonlinearBeamModelPostbuckling}. Other researchers have chosen a more numerically oriented solution to the same issue \cite{olesen-byskov_82a_asymptoticPostbucklingStresses, byskov_89a_smoothPostbucklingStresses, poulsen-damkilde_98a_directDetermination}. Another delicate point is the need to solve a nonlinear eigenvalue problem, which may be required in situations when the fundamental equilibrium path already shows significant nonlinearities and large bending deformations \cite{casciaro-etal_92a_asymptoticPostBuckling}. The nonlinear eigenvalue problem can be challenging for current solvers \cite{mehrmann-voss_04a_nonlinearEigenvalueChallenge}, even if fast and robust algorithms tailored to structural applications are available \cite{casciaro_06a_iterativeNonlinearEigenvalue}.

If all above points are carefully addressed, the asymptotic method gives a very accurate representation of the initial post-buckling response \cite{brezzi-etal_86a_getAroundQuadraticFold}, which can be further enhanced by, e.g., using Padé approximations \cite{cochelin-etal_94a_asymptoticNumericalPade, vannucci-etal_98a_asymptoticNumericalBranches, boutyour-etal_04a_asymptoticNumericalPade}. Moreover, the ANM can be used, with minor extra computational effort, to investigate the structures' imperfection sensitivity \cite{salerno-uva_06a_hoTheoremInteraction, salerno-23a_koitersWorstImperfection}.

The use of asymptotic post-buckling theory within structural optimization is still limited, and we are not aware of any investigations within TO. A few works are based on the optimization of post-buckling coefficients for composite design \cite{henrichsen-etal_16a_postBucklingKoiter}. In this work we apply a method based on the asymptotic post-buckling theory, to the sizing and topology optimization of truss and frame assemblies using the ground structure approach \cite{rozvany-etal_95a_layoutOptimizationStructures, gilbert-tyas_03a_optimizationPinJointedTrusses}. We show how constraining the lowest asymptotic coefficients effectively reduces imperfection sensitivity, and enhances the systems' initial post-buckling response. Moreover, the ANM obtained with the perturbation approach approximates the systems' nonlinear response, which is then used in the optimization. While unfolding the potentials of the method, we also highlight its difficulties, which are emphasized by referring to simple structural models composed of rods and beams assemblies (i.e., trusses and frames).

The paper is organized as follows. In \autoref{Sec:FormulationAsymptoticPostBuckling} we recall the essentials of the asymptotic post-buckling theory, showing its application to the analysis of simple structures. The discussion is kept compact, though we provide further computational details in \ref{App:DetailsFEM_2}-\ref{App:DetailsFEM}. The optimization problems considered are stated in \autoref{Sec:OptimizationProblemFormulation}, and we provide numerical examples, involving 2D truss and frame structures in \autoref{Sec:NumericalExamplesLarger}. \autoref{Sec:DiscussionOutlook} summarizes potentials and current challenges of the method, and its potential application to continuum, large-scale TO.

\begin{figure}[t]
 \centering
  \subfloat[]{
  \includegraphics[scale = 0.325, keepaspectratio]
  {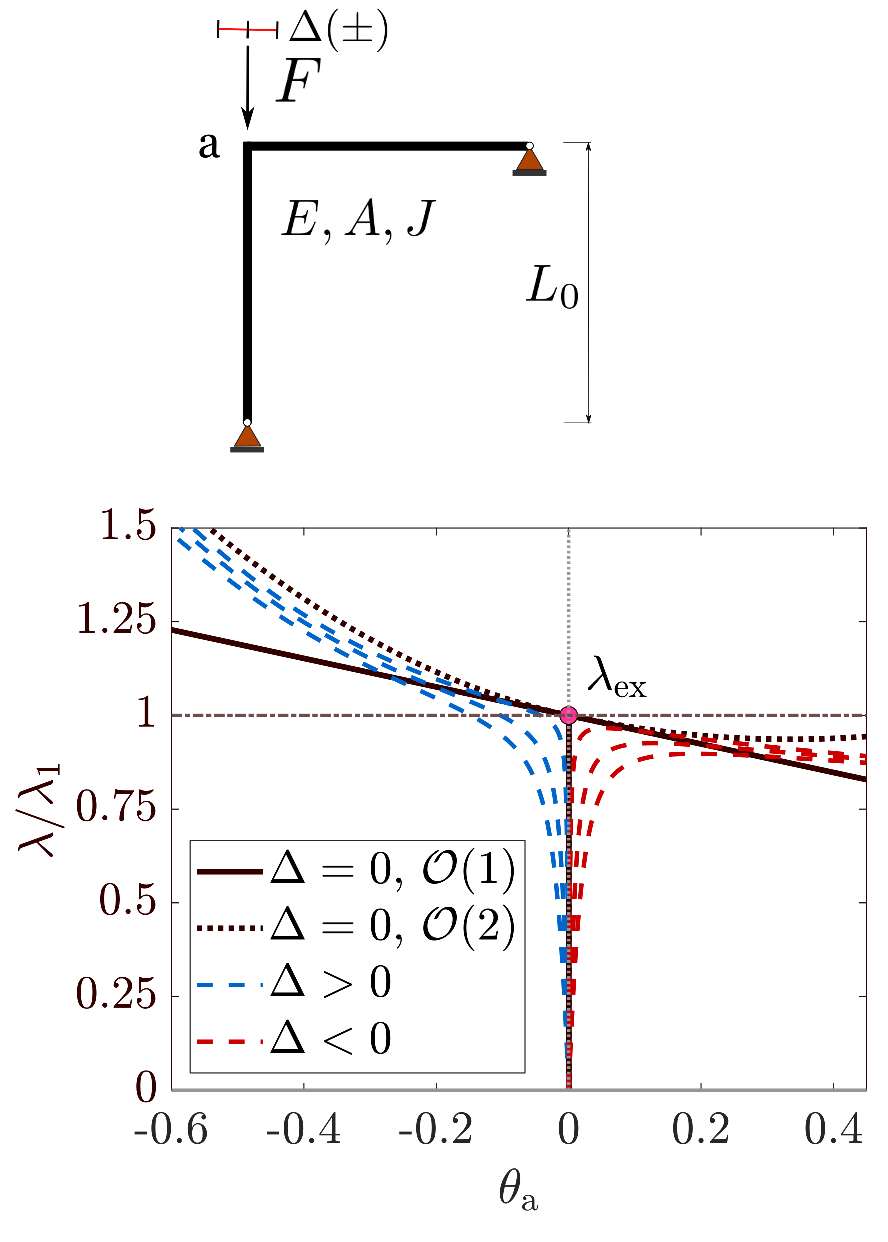}}
  \subfloat[]{
  \includegraphics[scale = 0.325, keepaspectratio]
  {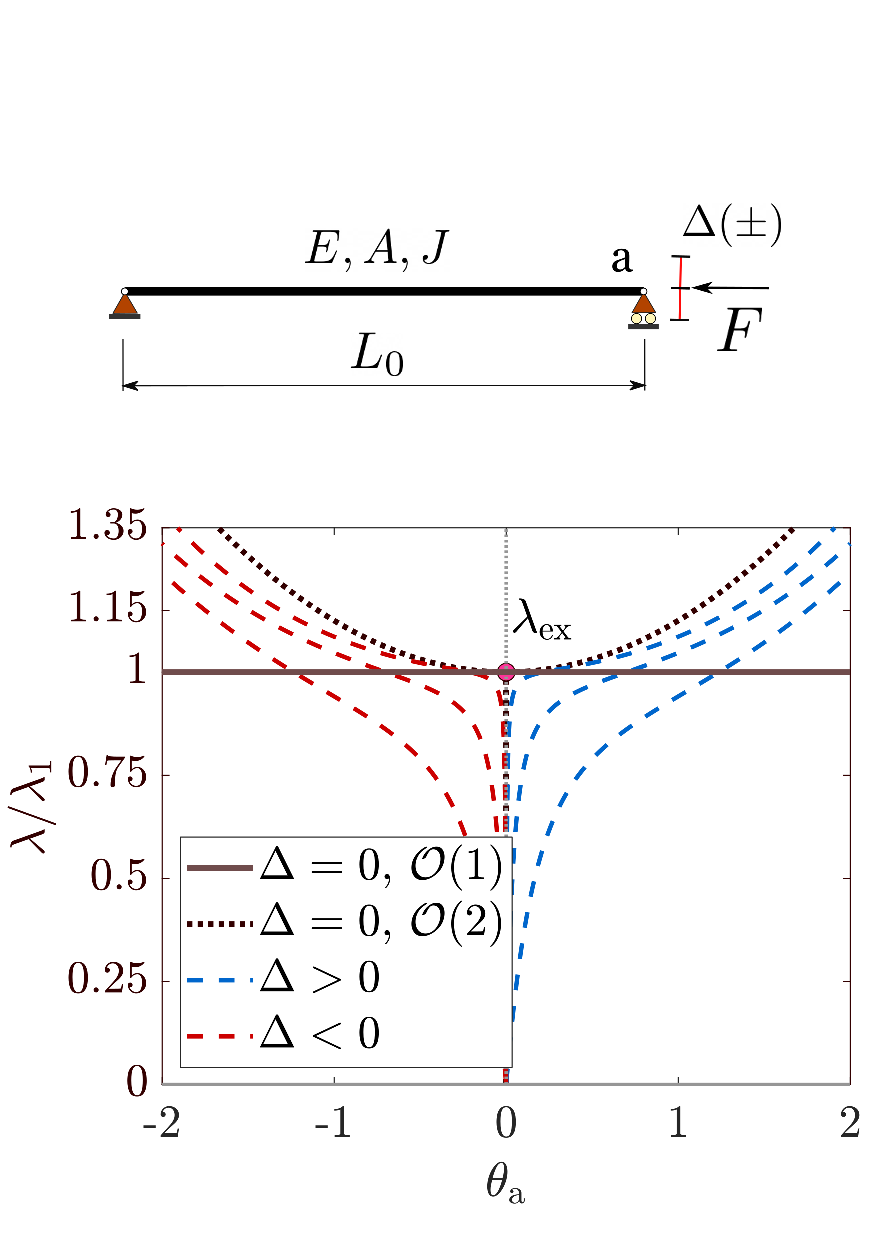}}
  \subfloat[]{
  \includegraphics[scale = 0.325, keepaspectratio]
  {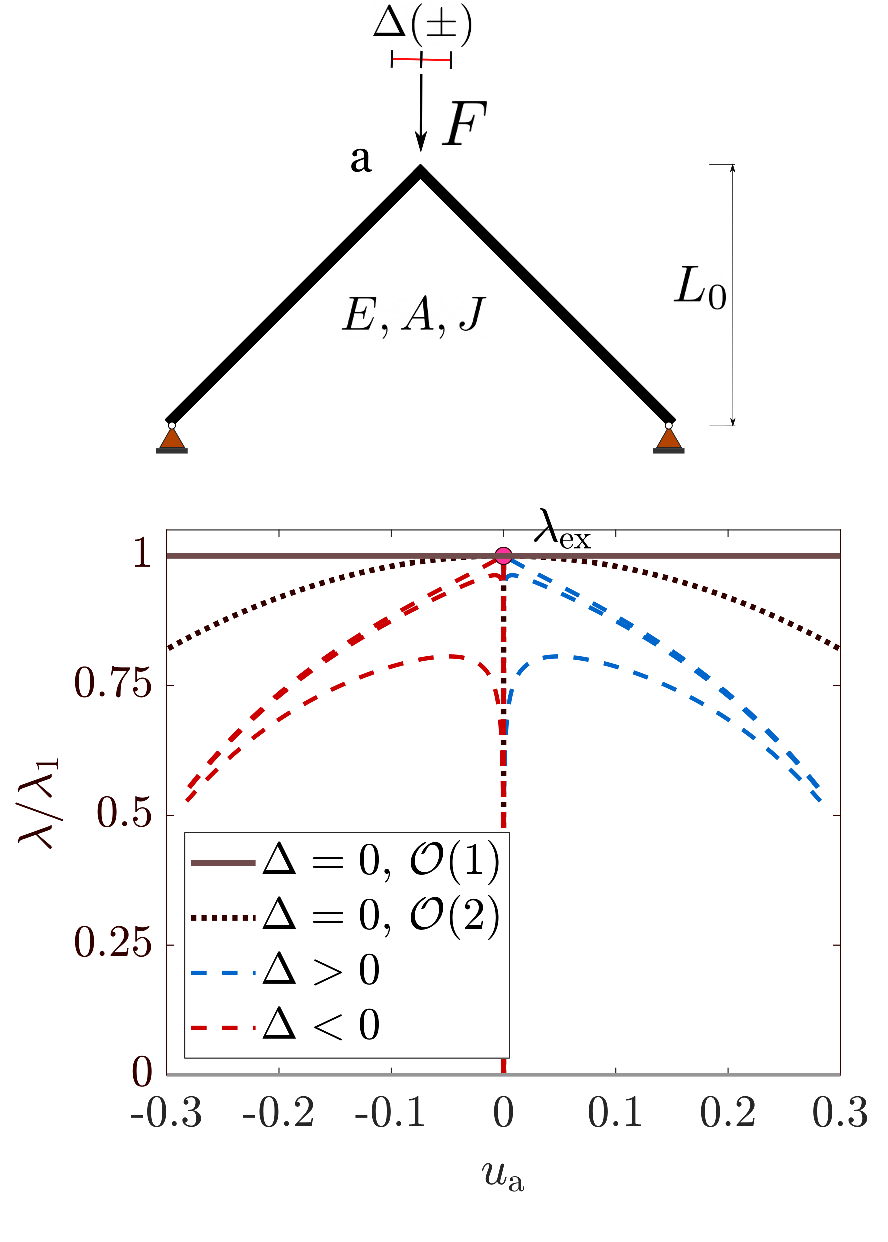}} 
  \caption{Types of bifurcations according to the signs of the post-buckling slope $\alpha$, and curvature $\beta$. (a) \emph{Asymmetric} bifurcation ($\alpha < 0$), for the Rooda-Koiter frame \cite{rooda-chilver_70a_frameBucklingPerturbation, olesen-byskov_82a_asymptoticPostbucklingStresses}. (b, c) \emph{Symmetric} bifurcations ($\alpha = 0$) for the Euler beam (b) and von Mises truss (c) \cite{book:bazant-cedolin2010}. For the last two cases, the bifurcations are respectively stable ($\beta > 0$) and unstable ($\beta < 0$), independent of the sign of the initial imperfection $\Delta$. Multiple dashed red and blue curves correspond to increasing magnitudes of the imperfection $\Delta$}
 \label{fig:CharacterBifurcation}
\end{figure}

\section{Essential equations of the asymptotic post-buckling theory}
 \label{Sec:FormulationAsymptoticPostBuckling}

We consider a discretized system with $n$ degrees-of-freedom (DOFs), acted upon by conservative loads whose distribution is described by the vector $\mathbf{f}_{0}\in\mathbb{R}^{n}$, and applied proportionally through the parameter $\lambda \in \mathbb{R}$ (i.e., $\mathbf{f}(\lambda) = \lambda\mathbf{f}_{0}$). We assume the existence of a fundamental equilibrium path, described by the curve $(\mathbf{d}(\lambda), \lambda)\in\mathbb{R}^{n}\times\mathbb{R}$, where $\mathbf{d}(\lambda)$ is the systems' equilibrium displacement.

Given a suitable abscissa $\xi\in\mathbb{R}$, the equilibrium path emerging from a simple bifurcation with buckling mode and load $(\mathbf{v}_{1}, \lambda_{1})$ can be described by the asymptotic expressions
\begin{equation}
 \label{eq:parametricExpression_DLambda}
  \begin{aligned}
   \mathbf{d}(\xi) & = \mathbf{u}_{0}(\lambda(\xi)) + \xi \mathbf{v}_{1} + \frac{\xi^{2}}{2}\mathbf{w} + \mathcal{O}(\xi^{3}) \\
   \lambda(\xi) & = \lambda_{1}\left(
   -\Delta\xi^{-1} + 
   1 + \alpha\xi + \beta\xi^{2} + 
   \mathcal{O}(\xi^{3})
   \right)
  \end{aligned}
\end{equation}
where $\Delta$ is the magnitude of the generic imperfection, whereas $\mathbf{u}_{0}$, $\mathbf{v}_{1}$ and $\mathbf{w}$ are the prebuckling displacement, buckling mode, and second order displacement correction, respectively.

These fields are computed from the following boundary value problems, obtained by perturbing the equilibrium equations at the bifurcation point \cite{koiter_45a_stabilityElasticEquilibrium, budiansky_74a_theoryBucklingPostBuckling}
\begin{align}
 \label{eq:zeroOrder_equilibrium}
  \delta\mathbf{d} \cdot \left(
  \Phi^{({\rm 1})}_{1} - \lambda_{1}\mathbf{f}_{0} \right) & = 0
  \:, \qquad \forall\:\delta\mathbf{d} \\
  \label{eq:firstOrder_eigenvalueEquation}
  \delta\mathbf{d} \cdot 
  \Phi^{({\rm 2})}_{1}\mathbf{v}_{1} & = 0
  \:, \qquad \forall\:\delta\mathbf{d} \: , \: \mathbf{v}_{1}\neq 0 \\
  \label{eq:secondOrderDisplacementCorrection}
  \delta\mathbf{d} \cdot \left(
  \Phi^{({\rm 2})}_{1}\mathbf{w} + \frac{1}{2}\Phi^{({\rm 3})}_{1}\mathbf{v}^{2}_{1} + 
  \alpha\lambda_{1}\Phi^{({\rm 3})}_{1}\mathbf{u}_{0}\mathbf{v}_{1} \right) & = 0
  \: , \qquad \forall\:\delta\mathbf{d} \: ,  \: \mathbf{w}\cdot \mathbf{v}_{1} = 0
\end{align}
where $\Phi^{(n)}_{1} := [\nabla^{(n)}_{\mathbf{d}}\Phi(\mathbf{u}_{0}(\lambda), \lambda)]_{\lambda = \lambda_{1}}$ is the $n$-th order strain energy variation, evaluated at the bifurcation point (see \ref{App:DetailsFEM} for details). \autoref{eq:zeroOrder_equilibrium} and \eqref{eq:firstOrder_eigenvalueEquation} are the (generally nonlinear) pre-buckling equilibrium equation and buckling eigenvalue problem, respectively.\autoref{eq:secondOrderDisplacementCorrection} is a higher order equation providing a unique solution for the second order displacement correction $\mathbf{w}$, once paired with the orthogonality condition $\mathbf{w}\cdot\mathbf{v}_{1} = 0$.

The coefficients $\alpha$ and $\beta$ in \eqref{eq:parametricExpression_DLambda} describe the initial slope and curvature of the post-buckling path, and are defined as follows \cite{budiansky_74a_theoryBucklingPostBuckling}
\begin{equation}
 \label{eq:asymptoticCoefficients}
  \begin{aligned}
   \alpha & = -\frac{1}{2\lambda_{1}}
   \Phi^{({\rm 3})}_{1}\mathbf{v}^{3}_{1} \\
   \beta  & = -\frac{1}{\lambda_{1}} 
   \left[
   \frac{1}{6}
   \Phi^{({\rm 4})}_{1}\mathbf{v}^{4}_{1} + 
   \Phi^{({\rm 3})}_{1}\mathbf{w}\mathbf{v}^{2}_{1} +
   \alpha\lambda_{1}
   \left( \Phi^{({\rm 3})}_{1}\mathbf{u}_{0}\mathbf{v}_{1}\mathbf{w} + \frac{1}{2} \Phi^{({\rm 4})}_{1}\mathbf{u}_{0}\mathbf{v}^{3}_{1} \right) + \frac{\alpha^{2}\lambda^{2}_{1}}{2}\Phi^{({\rm 4})}_{1}\mathbf{u}^{2}_{0}\mathbf{v}^{2}_{1}
   \right]
  \end{aligned}
\end{equation}
where we assumed the normalization $(\Phi^{({\rm 3})}\mathbf{u}_{0}(\lambda))_{\lambda = \lambda_{1}}\mathbf{v}^{2}_{1} = 1$.

Some details on the finite element (FE) formulation of \eqref{eq:parametricExpression_DLambda}-\eqref{eq:asymptoticCoefficients}, and on the associated computational cost are given in \ref{App:DetailsFEM_1}, and the extension to the case of multiple buckling modes, either simultaneous or separate is outlined in \ref{App:DetailsFEM_2}.

The asymptotic coefficients \eqref{eq:asymptoticCoefficients} determine the character of the bifurcation point $(\mathbf{v}_{1}, \lambda_{1})$ associated with the \emph{perfect} structure, and thus also the response when considering imperfections \cite{koiter_45a_stabilityElasticEquilibrium, budiansky_74a_theoryBucklingPostBuckling}. Indeed, the actual response and value of the critical load, say $\lambda_{c}$, will depend on the loading and geometrical imperfections (hereafter generically denoted by $\Delta$) assigned to the structure.

In particular, $\alpha \neq 0$ corresponds to an \emph{asymmetric} bifurcation, leading to a so-called ``snap-buckling'' phenomenon when paired with an imperfection of opposite sign (thus, $\alpha\Delta < 0$). For $\alpha = 0$ we have a \emph{symmetric} bifurcation, and its post-buckling stability is determined by the sign of $\beta$. The bifurcation is always stable for $\beta > 0$, and always unstable for $\beta < 0$, independent of the sign of the imperfection $\Delta$.

These three types of response are illustrated in \autoref{fig:CharacterBifurcation}, referring to the Rooda-Koiter frame \cite{rooda-chilver_70a_frameBucklingPerturbation}, Euler beam \cite{pignataro-etal_82a_nonlinearBeamModelPostbuckling}, and von Mises frame \cite{book:bazant-cedolin2010}, respectively. The Rooda-Koiter frame shows an asymmetric bifurcation, and the post-buckling slope $\alpha$ is sufficient to describe its initial post-buckling response. The curvature $\beta$ carries additional information about the post-buckling load stiffening (see dotted black line in \autoref{fig:CharacterBifurcation}(a)). For both the other examples we have $\alpha = 0$. Therefore, to determine their initial post-buckling response we need to compute $\beta$.

Other than describing the initial post-buckling response, the asymptotic coefficients $\alpha$ and $\beta$ in \eqref{eq:parametricExpression_DLambda} allow the approximation of the equilibrium path $(\mathbf{d}(\xi), \lambda({\xi}))$ at a given abscissa $\xi$, weighting the contribution of the buckling mode $\mathbf{v}_{1}$ in the total displacement $\mathbf{d}(\xi)$. For a first-order approximation we only need the post-buckling slope $\alpha$, and thus quantities already computed by the eigenvalue buckling analysis, \eqref{eq:zeroOrder_equilibrium} and \eqref{eq:firstOrder_eigenvalueEquation}. 

On the other hand, to build a second-order approximation we also need to compute $\beta$, which requires solving \eqref{eq:secondOrderDisplacementCorrection} for the second order displacement correction $\mathbf{w}$. This latter step requires some caution as $\beta$ can be accurately determined only if the fourth-order energy variation appearing in both \eqref{eq:secondOrderDisplacementCorrection} and \eqref{eq:asymptoticCoefficients} is free from spurious effects.

This can be achieved by using geometrically exact beam models, free from spurious deformations \cite{book:antman, pignataro-etal_82a_nonlinearBeamModelPostbuckling}, and by properly treating the axial and transversal displacement components, avoiding locking phenomena \cite{NOTE_garcea-etal_13a_postBucklingAsymptotic, olesen-byskov_82a_asymptoticPostbucklingStresses, byskov_89a_smoothPostbucklingStresses}. The rod and beam models used in this work are shortly described in \ref{App:DetailsFEM}, and we refer to \cite{salerno-lanzo_97a_nonlinearFElockingFree, garcea-etal_12a_implicitCoRotationalModel} for further details.

\subsection{Application to simple test cases}
 \label{sSec:AsymptoticPBanalysisSimpleCases}

Before moving to the optimization applications, we test the accuracy of the asymptotic-numerical method (ANM) outlined above for the analysis of simple structures. These are modeled by beams (see \ref{sApp:beamElement&discretization} for details); thus, the generalized displacement vector contains axial, transversal and rotational DOFs. In all the following examples, the beam length, Young's modulus, and cross section properties are $L_{0} = 1$ m, $E = 21$ MPa (rubber-like material), $A = 10^{-2} $ m$^{2}$, and $J = 10^{-4}/12$ m$^{4}$, respectively.

\begin{figure}[t]
 \centering
  \subfloat[Euler beam (cfr. \autoref{fig:CharacterBifurcation}(b))]{
  \includegraphics[scale = 0.4, keepaspectratio]
  {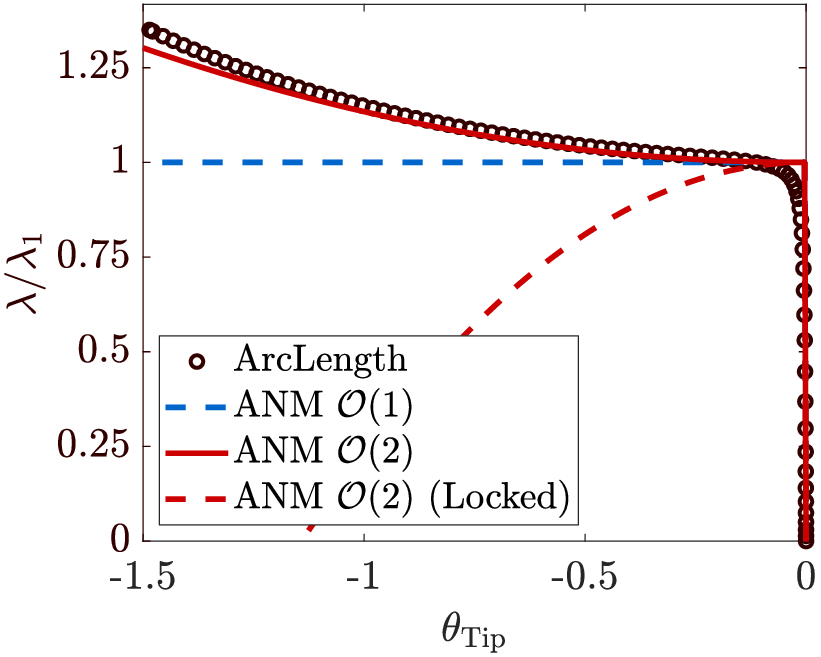}} \qquad
  \subfloat[Rooda-Koiter frame (cfr. \autoref{fig:CharacterBifurcation}(a))]{
  \includegraphics[scale = 0.4, keepaspectratio]
  {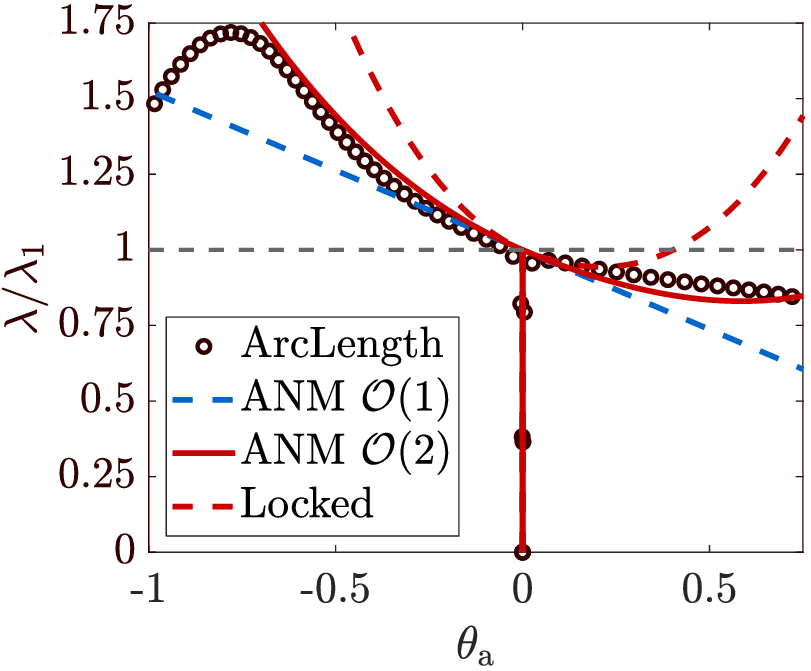}} \\
  \subfloat[Rooda-Koiter assembly modeled with $m=704$ beam elements]{
  \includegraphics[scale = 0.4, keepaspectratio]
  {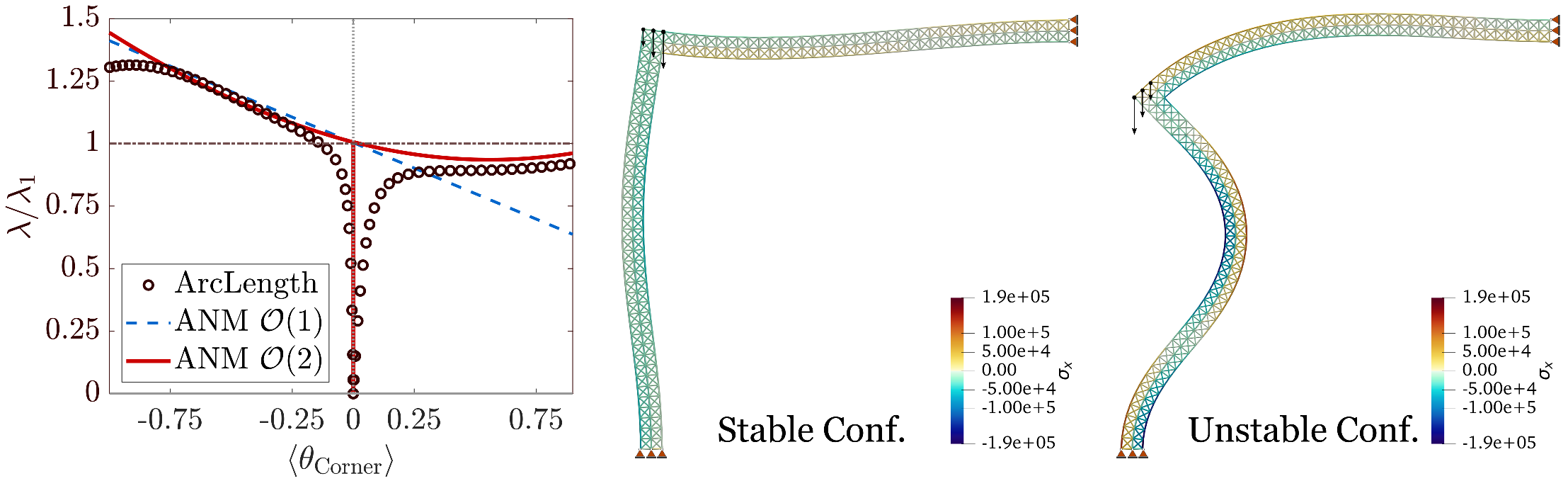}}
  \caption{Equilibrium paths computed by the arc-length method (black circles), compared to the first-order (blue dashed curve) and second-order (red continuous curve) approximations given by the ANM method. The red dashed lines in (a, b) correspond to the second-order ``locked'' asymptotic solution, following from inappropriate structural modeling and discretization. The plot in (c) also shows the nonlinear deformation at the buckling point, for the stable and unstable Rooda-Koiter configuration, respectively}
 \label{fig:AsymptoticVerificationLiteratureTest}
\end{figure}

\paragraph{Euler beam} We refer to the configuration sketched in \autoref{fig:CharacterBifurcation} (b), where the force at the right end is $F = -10^{3} N$, and the load misalignment $\Delta = -10^{-3}$ generates a clock-wise couple. The closed-form solution, when considering large deformations, is given by the \emph{elastica} theory \cite{book:bazant-cedolin2010}, and it features $\lambda_{1} = \frac{\pi^{2}EJ}{F L^{2}_{0}} \approx 1.7272$, $\alpha = 0$, and $\beta = \lambda_{1}/4$.

The numerical solution, computed with $m = 10$ beam elements based on the Green-Lagrange strain measure (see \ref{sApp:beamElement&discretization}), correctly recovers the pre-buckling displacement, buckling mode and load factor, and also the first post-buckling coefficient $\alpha = 0$. However, the post-buckling curvature is clearly incorrect ($\beta \approx -0.89\lambda_{1}$), leading to a ``locking'' of the post-buckling response, (see dashed red line in \autoref{fig:AsymptoticVerificationLiteratureTest} (a)).

This error is not entirely due to discretization, but is embodied in the structural modeling itself. Thus, we cannot expect it to vanish with mesh refinement. Indeed, the \emph{closed-form} solution to \eqref{eq:zeroOrder_equilibrium}-\eqref{eq:secondOrderDisplacementCorrection}, when using the Green-Lagrange strains, reads \cite{NOTE_casciaro_05a_computationalPost-buckling}
\begin{equation}
 \mathbf{u}_{0}(\lambda) =
 \left\{ \begin{array}{c}
 -\frac{\lambda F_{0}}{EA} x \\ 0 \\ 0 \end{array} \right\} \ , \ \
 \mathbf{v}_{1} =
 \left\{ \begin{array}{c}
 0 \\ \frac{a_{1}L_{0}}{\pi}\sin(\frac{\pi x}{L_{0}}) \\ 
 a_{1}\cos(\frac{\pi x}{L_{0}}) \end{array} \right\} \ , \ \ 
 \mathbf{w} =
 \left\{ \begin{array}{c}
 a_{1}(\frac{x}{2} + \frac{\sin(2\pi x)}{4\pi}) \\ 0 \\ 0 \end{array} \right\}
\end{equation}
where the second order correction $\mathbf{w}$ induces an axial strain field, not present in the elastica theory. As a consequence, the post-buckling curvature takes the wrong value $\beta = -3/4\lambda_{1}$.

To avoid spurious strains in the high order deformation field, we refer to the beam model proposed in \cite{garcea-etal_12a_implicitCoRotationalModel} (see \ref{App:DetailsFEM}), allowing to compute the correct curvature and to achieve a high accuracy approximation of the post-buckling displacement, see red continuous line in \autoref{fig:AsymptoticVerificationLiteratureTest} (a).

We remark that the correct post-buckling slope $\alpha = 0$ is also computed when using the Green-Lagrange strain. However, this would give little information for this case, as the post-buckling stiffening would not be represented.

\paragraph{Rooda-Koiter frame} We refer to the configuration of \autoref{fig:CharacterBifurcation} (a), where the downward force $F = -10^{3}$ N is applied at the corner node. The imperfection $\Delta = \pm 10^{-3}$, again generating a couple, is now considered with both positive and negative signs, to trigger the asymmetric post-buckling response. For the chosen set of data, the closed form solution gives $\lambda_{1} = 2.430$, $\alpha = 0.3805$ and $\beta = 0.1421$ \cite{book:bazant-cedolin2010}. \autoref{fig:AsymptoticVerificationLiteratureTest} (b) shows the arc-length solution as circular marks, obtained by tracking the rotation at the corner node.

The ANM gives a high quality approximation of the initial post-buckling path. The numerically computed coefficients are $\lambda_{1} = 2.4313$, $\alpha = 0.3820$, and $\beta = 0.1436$, with relative errors of about $0.05\%$, $0.01\%$ and $0.1\%$, respectively. We highlight that the post-buckling slope can also be computed based on the simple Green-Lagrange strain, with a negligible error ($\approx 0.2\%$). However, the wrong evaluation of the second-order coefficient would once again lead to locking of the post-buckling response, as shown by the dashed line in \autoref{fig:AsymptoticVerificationLiteratureTest} (b).

\paragraph{Larger truss and frame assemblies}
The ANM is also tested on structural assemblies, made up of hundreds or thousands of bars, either modeled as rods or beams. As an example, \autoref{fig:AsymptoticVerificationLiteratureTest} (c) shows a Rooda-Koiter frame, modeled with $m = 704$ bars, each featuring the material and cross section properties listed at the top of this section. The side length of the overall square domain is $L_{0} = 1$, and the vertical and horizontal members have width $t = 0.063$, and contain three rows of beams, with inner horizontal and crossing (non-connected) members. The nodes at the end sections of the members are hinged (thus $u = v = 0$), and the load is applied at the three leftmost nodes at the top. To mimic the imperfection with the $\pm$ signs, we consider two load cases, with the load magnitude linearly increasing or decreasing across the three nodes.

Even if we cannot refer to a closed-form solution for this case, the asymptotic coefficients are in line with the values from the simpler beam model discussed before ($\alpha = 0.3104$ and $\beta = 0.2501$). Moreover, we can see that the accuracies obtained by the linear $\mathcal{O}(1)$ and quadratic $\mathcal{O}(2)$ ANM closely match the results obtained by the path-following algorithm, especially for the branch corresponding to the stable configuration (increasing post-buckling load).

Compared to the previous two examples, this model shows a higher imperfection sensitivity. Indeed, we observe larger pre-buckling deformations, and now the maximum load multiplier attained on the ``right'' branch (the one corresponding to the unstable configuration), is sensibly lower than the BLF of the perfect structure. This is expected, given the more complex modeling, with many bars, and the mixed bending-compression stress state.

\section{Application of the ANM to sizing and topology optimization}
 \label{Sec:OptimizationProblemFormulation}

We consider an optimization problem with the following general statement, for now without accounting for the initial post-buckling coefficients
\begin{equation}
 \label{eq:optProblemYesStabilityConstraints}
  \begin{cases}
   \min\limits_{\mathbf{x}\in [0,1]^{m}} &
   W(\mathbf{x}, \mathbf{d}(\mathbf{x},\lambda)) \\
   {\rm s.t.}
   & \mathcal{R}(\mathbf{d}(\mathbf{x}, \lambda)) = 0 \\
   & g_{      V} := v_{f}(\mathbf{x})/\bar{v}_{f} - 1 \leq 0 \\
   & g_{\lambda} := 1 - \lambda_{1} / \lambda_{\rm ref} \leq 0
  \end{cases}
\end{equation}
where $g_{V}$ and $g_{\lambda}$ are the volume and buckling constraints, respectively.

The first refers to the prescribed volume fraction $\bar{v}_{f}$, whereas the second is based on a reference load factor $\lambda_{\rm ref}$, specified in the following for each example, and is meant to impose global stability \cite{weldeyesus-etal_20a_trussTopologyOptimizationStability}. The buckling load $\lambda_{1}$, and displacement $\mathbf{v}_{1}$, are computed either via a Linearized Buckling Analysis (LBA), or a nonlinear buckling analysis (NLBA), depending on the influence of nonlinearities in the pre-buckling regime (see discussion in \autoref{Sec:DiscussionOutlook} and \ref{App:DetailsFEM_1}).

To control the initial post-buckling response, either or both the following constraints can be introduced in the optimization problem \eqref{eq:optProblemYesStabilityConstraints}
\begin{equation}
 \label{eq:constraintAlphaBeta}
  g_{\alpha} = |\alpha| \leq \bar{\alpha} \: , \qquad
  g_{\beta} = \beta \geq \underline{\beta}
\end{equation}

In particular, $g_{\alpha}$ controls the post-buckling slope, avoiding a structural configuration with asymmetric (and thus unstable) post-buckling response. However, this may not be sufficient to avoid imperfection sensitivity and an initial drop in the post-buckling load (see \autoref{fig:CharacterBifurcation} (c)). Thus, we also need $g_{\beta}$ to impose a minimum positive curvature $\underline{\beta}$.

The objective in \eqref{eq:optProblemYesStabilityConstraints} may be chosen as the linear compliance $W_{0} = \mathbf{f}^{T}_{0}\mathbf{u}_{0}$, where the force and displacement vectors are linked by the linear equation $\mathcal{R}(\mathbf{u}_{0}) = \mathbf{f}_{0} - K_{0}(\mathbf{x})\mathbf{u}_{0}$, and $K_{0}$ is the linear part of the tangent matrix. Alternatively, we may choose one of the following nonlinear response functions: end-compliance ($W_{\rm EC}$), or the complementary energy ($W_{\rm CE}$) \cite{buhl-etal_00a_geometricNonlinearTO, bruns-etal_2002a_numericalSnapThrough, kemmler-etal_05a_largeDeformationStability}
\begin{equation}
 \label{eq:nonLinearResponses_ECCW}
 W_{EC} = \bar{\lambda}\mathbf{f}^{T}_{0}\mathbf{d}(\mathbf{x},\bar{\lambda}) \: , \qquad
 W_{CE} = \int^{\bar{\lambda}}_{0}
 \mathbf{f}(\lambda)^{T}\mathbf{d}(\mathbf{x},\lambda) \: {\rm d}\lambda
\end{equation}
where $\bar{\lambda}$ is the prescribed load value, which may be below or above the critical one.

In the nonlinear case, the equilibrium equation reads
\begin{equation}
 \label{eq:stateEquation_nonlinear}
  \mathcal{R}(\mathbf{d}(\mathbf{x}, \lambda)) = \lambda\mathbf{f}_{0} - \mathbf{f}_{i}(\mathbf{x}, \mathbf{d}(\mathbf{x},\lambda))
\end{equation}
where $\mathbf{f}_{i}(\mathbf{x},\mathbf{u}(\mathbf{x},\lambda))$ is the internal forces vector and, in a ``classical'' nonlinear analysis, \eqref{eq:stateEquation_nonlinear} is solved for each equilibrium point, e.g. by a Newton iteration
\begin{equation}
 \label{eq:newtonStep}
 K_{t}(\mathbf{x},\mathbf{u}_{k}(\mathbf{x},\lambda_{k}))\Delta\mathbf{u}_{k} = -\mathcal{R}(\mathbf{u}_{k}(\mathbf{x}, \lambda_{k}))
\end{equation}
where $K_{t}$ is the tangent stiffness matrix. Then, a path-following strategy may be needed to obtain the complete equilibrium path, depending on the objective function selected \cite{buhl-etal_00a_geometricNonlinearTO, bruns-etal_2002a_numericalSnapThrough}.

In the following, we replace all the nonlinear incremental procedures by the ANM described in \autoref{Sec:FormulationAsymptoticPostBuckling}, when solving the optimization problems. We carry out a path-following solution of \eqref{eq:stateEquation_nonlinear} only for post-evaluating the design's response, for a given perturbation. In this post-evaluation step, we define the actual critical load, say $\lambda_{c}$, as the first load level at which $|K_{t}| \leq 0$.

\begin{figure}[t]
 \centering
  \subfloat[]
  {\includegraphics[scale = 0.425, keepaspectratio]
  {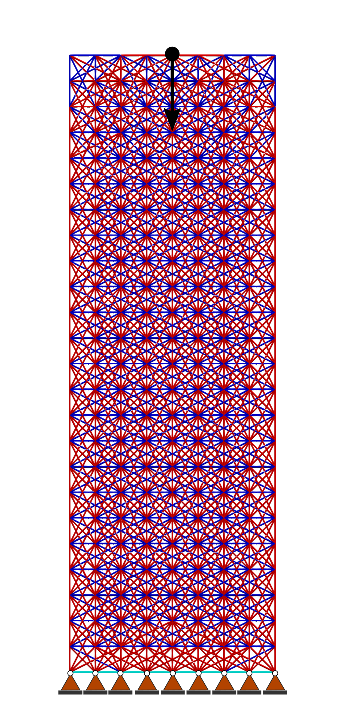}}
  \qquad
  \subfloat[]
  {\includegraphics[scale = 0.425, keepaspectratio]
  {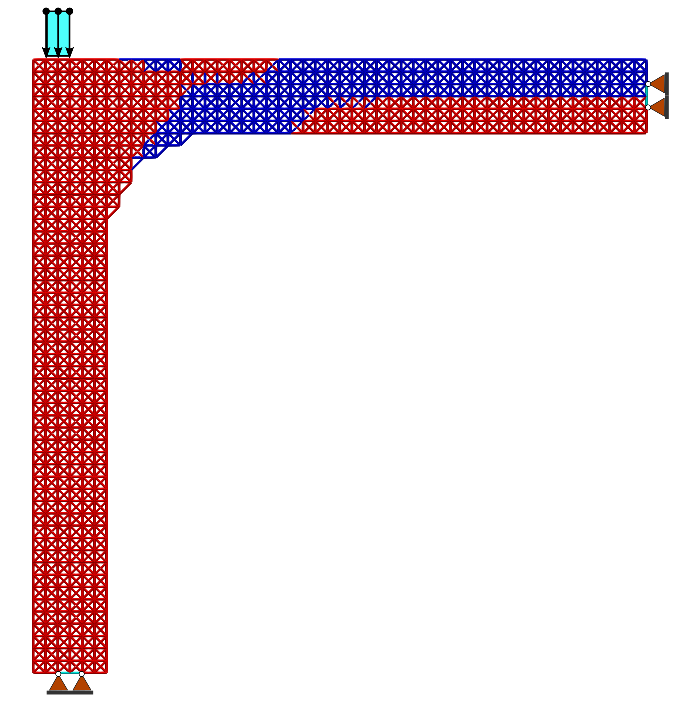}}
  \qquad
  \subfloat[]
  {\includegraphics[scale = 0.475, keepaspectratio]
  {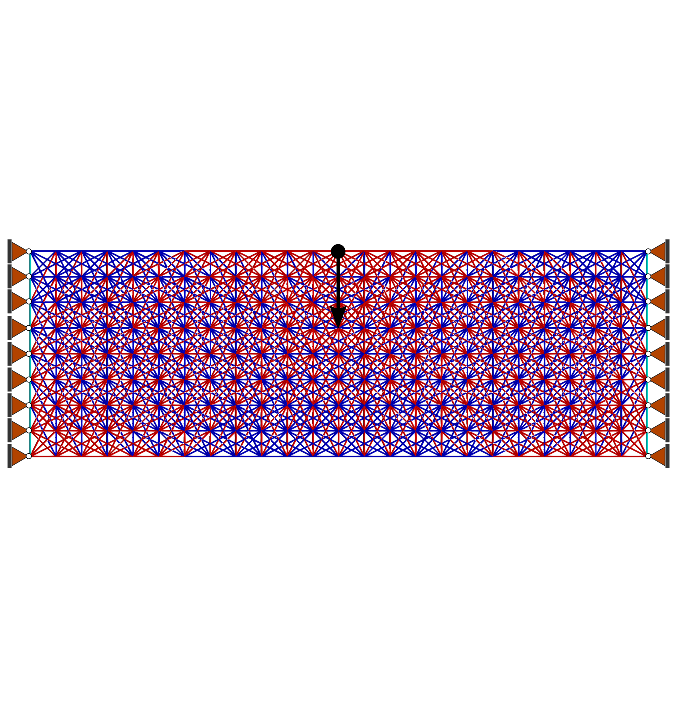}}
 \caption{Initial design (IG) for the compressed column (a), Rooda frame (b) and deep clamped beam (c) configurations, examined in \autoref{sSec:MotivationExample}-\autoref{sSec:ClampedDeepBeam}. Here and in the following plots, bars subjected to tensile axial stress are colored in blue, whereas those in red are compressed}
  \label{fig:NumericalExamplesInitialGuesses}
\end{figure}

\section{Numerical examples}
 \label{Sec:NumericalExamplesLarger}

In the following examples, the design variables $\mathbf{x} = \{x_{e}\}^{m}_{e=1}$ are used in a SIMP-like interpolation \cite{bendsoe-sigmund_99a_materialInterpolationSchemes} to describe the rods' and beams' Young modulus, as $E(x_{e}) = E_{\rm min} + x_{e}(E_{0} - E_{\rm min})$, where $E_{0}$ is the Young modulus of the solid material. For truss modeling (i.e., assembly of rods) we set $E_{\rm min} = 10^{-3}E_{0}$, as this relative high value helps stabilizing the optimization history when imposing high values of the minimum buckling constraint. Therefore, the optimization here is categorized as a reinforcement, or sizing, problem. When using frame modeling (assembly of beams), we set $E_{\rm min} = 10^{-9}E_{0}$, so here we are properly considering a TO problem.

Problem \eqref{eq:optProblemYesStabilityConstraints} is solved in a nested form, using the Method of Moving Asymptotes to update the design variables \cite{svanberg_87a}. The buckling constraint $g_{\lambda}$ is enforced through a bound formulation \cite{bendsoe-olhoff_85a_designVibrationsBeamsShafts}, introducing separate constraints on the lowest $q$ buckling load factors (BLFs).

The steps necessary for setting up the ANM, and computation of the post-buckling coefficients, are outlined in \ref{App:DetailsFEM_2} and \ref{App:DetailsFEM_1}, and entail the use of the rod and beam models described in \ref{App:DetailsFEM}. When post-evaluating the response of a given design, the ``high fidelity'' equilibrium path is obtained with a minimum-norm arc-length algorithm \cite{book:deborst2012}, using a co-rotational description for both rods and beams \cite{book:krenk2009}, and solving \eqref{eq:stateEquation_nonlinear}-\eqref{eq:newtonStep} at each point.

\begin{figure}[t]
 \centering
  \subfloat[]
  {\includegraphics[scale = 0.375, keepaspectratio]
  {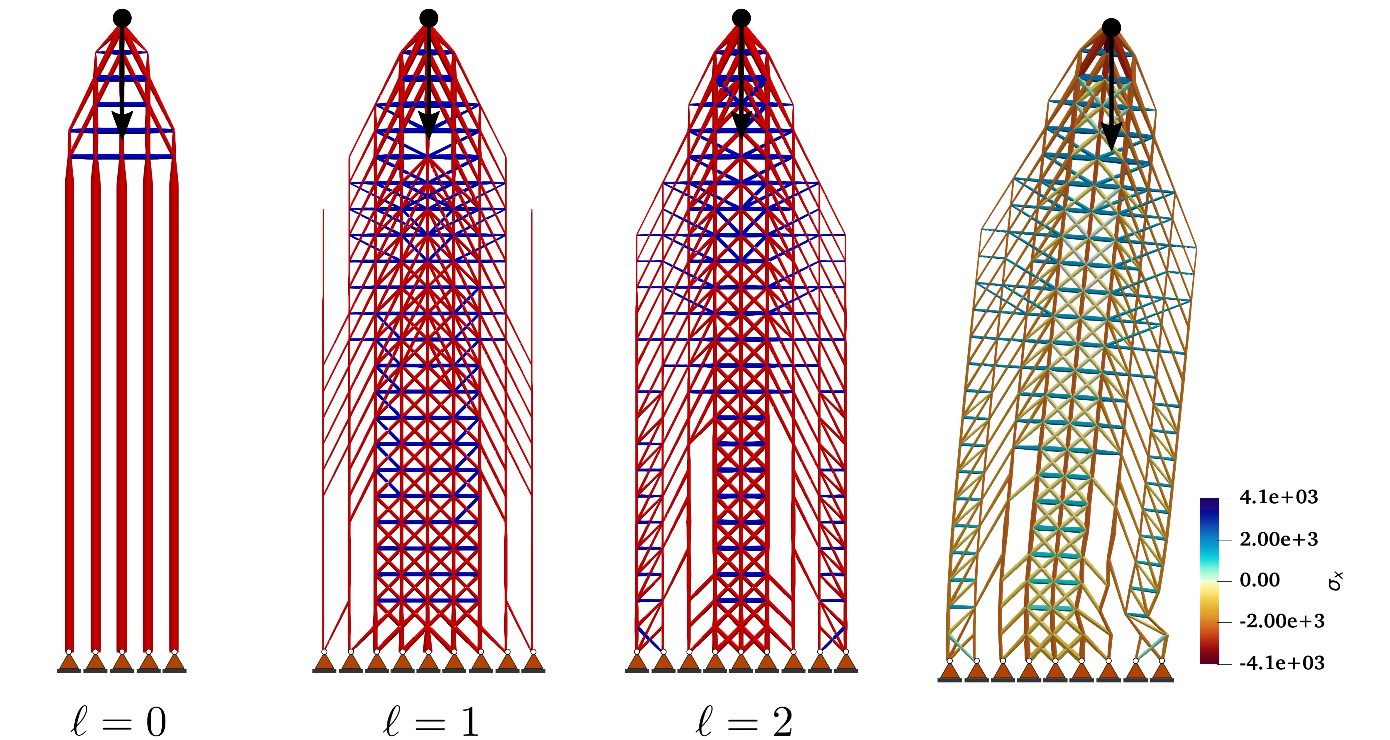}}
  \subfloat[]
  {\includegraphics[scale = 0.415, keepaspectratio]
  {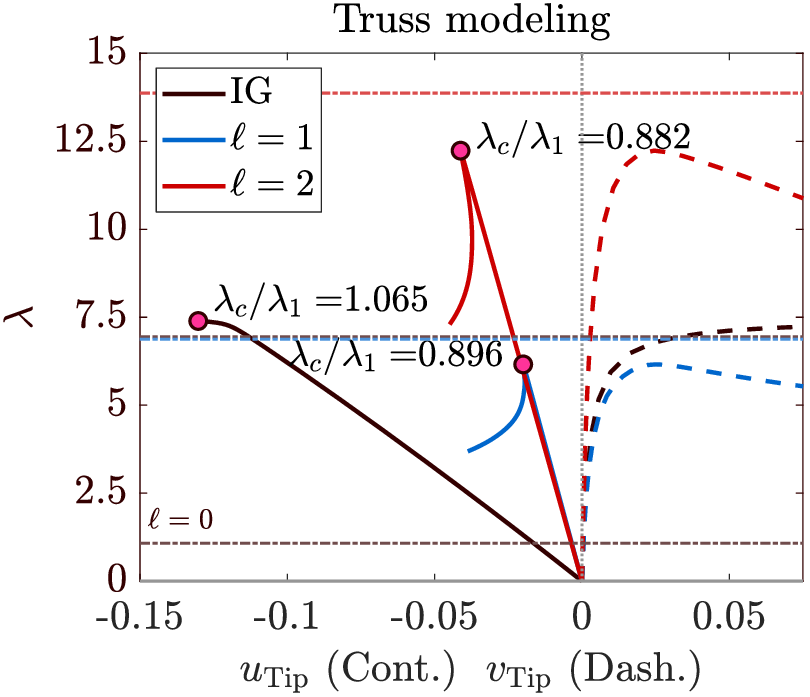}} \\
  \subfloat[]
  {\includegraphics[scale = 0.375, keepaspectratio]
  {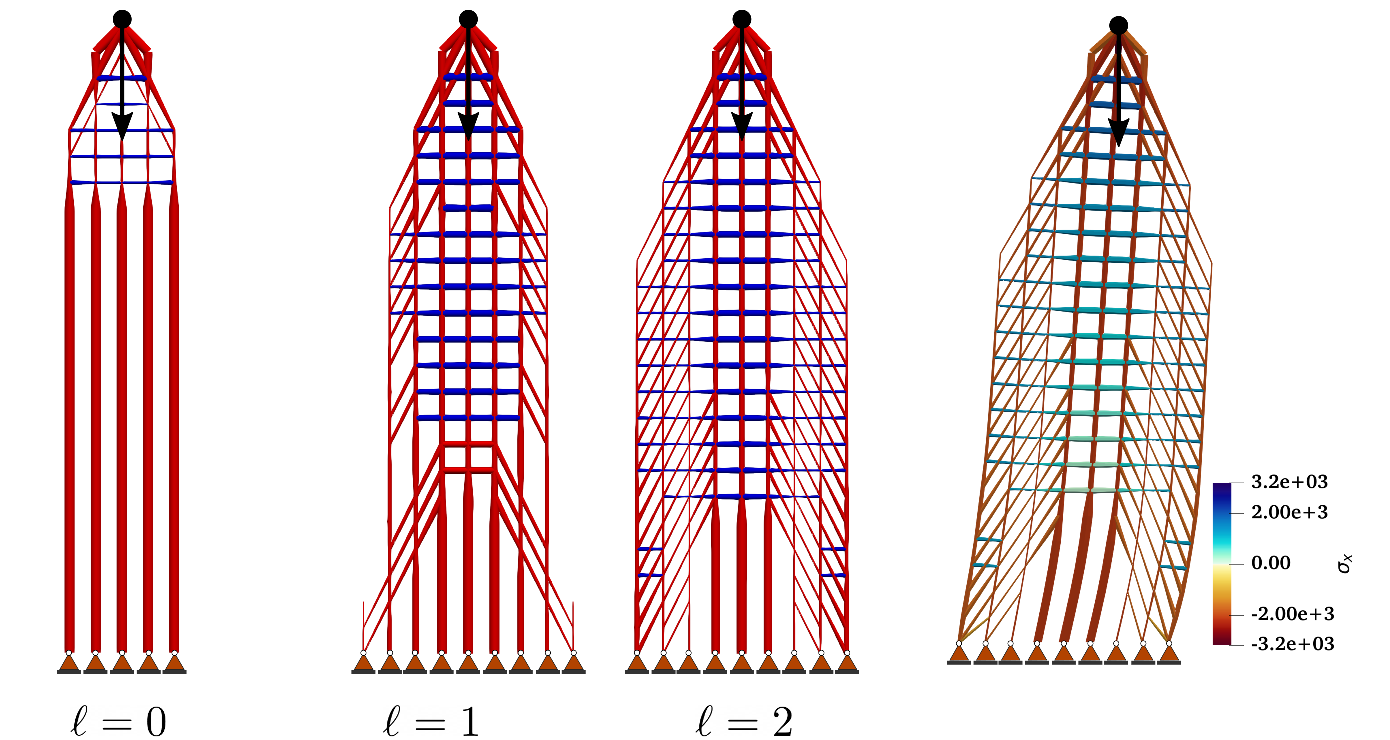}}
  \subfloat[]
  {\includegraphics[scale = 0.415, keepaspectratio]
  {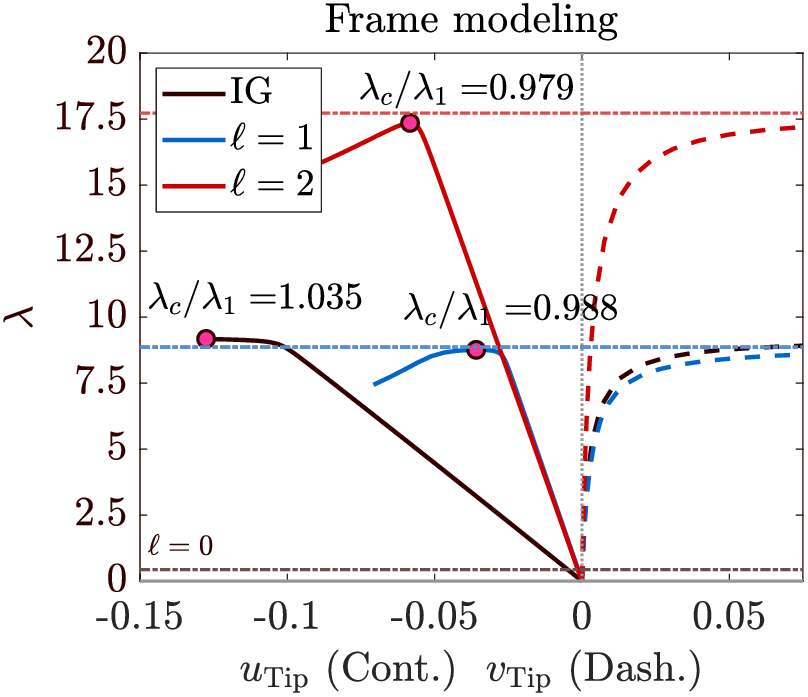}}
 \caption{Column designs obtained by linear compliance minimization, with volume and buckling constraints, considering truss (a) and frame (c) modeling. The volume fraction is $\bar{v}_{f} = 0.25$ and $\lambda_{\rm ref} = \ell\lambda_{1(0)}$, where $\lambda_{1(0)}$ is the BLF of the initial design (see \autoref{tab:Table1}). (b) and (d) show the nonlinear response of the structures, computed by the arc-length algorithm up to reaching $v_{\rm Tip} \leq 0.075$. The horizontal dash-dotted lines mark the BLFs ($\lambda_{1}$) computed by the LBA, whereas the red dots mark the maximum loads reached by the nonlinear response ($\lambda_{c}$)}
  \label{fig:MotivationExample_Column_optDesigns1}
\end{figure}

\subsection{Motivation example: design of a compressed column}
 \label{sSec:MotivationExample}

This example demonstrates how including the initial post-buckling slope and curvature in the optimization achieves a stable design, with reduced imperfection sensitivity.

We consider the compressed column with aspect ratio $3:1$ shown in \autoref{fig:NumericalExamplesInitialGuesses} (a), modeled either as a truss or a frame. In both cases, we consider a grid of $24\times 8$ squares with side length $h$, and we introduce bars with connectivity within the radius $r = 2\sqrt{2}h$, keeping only the shortest of collinear elements.

The design domain contains $m = 1,504$ bars, each featuring solid material and cross section properties $E_{0} = 21$ MPa, $A = 10^{-2}$ m$^{2}$, and $J = 10^{-4}/12$ m$^{4}$. The tip compressive force has magnitude $|F| = 10^{3}$ N, and the rightward pointing force, with magnitude $|\Delta F| = 10^{-3}|F|$, is applied as a source of perturbation when evaluating the systems' nonlinear response, not when performing the optimization. Both ($u,v$) displacements are restrained at the bottom edge.

\begin{figure}[t]
 \centering
 \subfloat[]{
  \includegraphics[scale = 0.375, keepaspectratio]
  {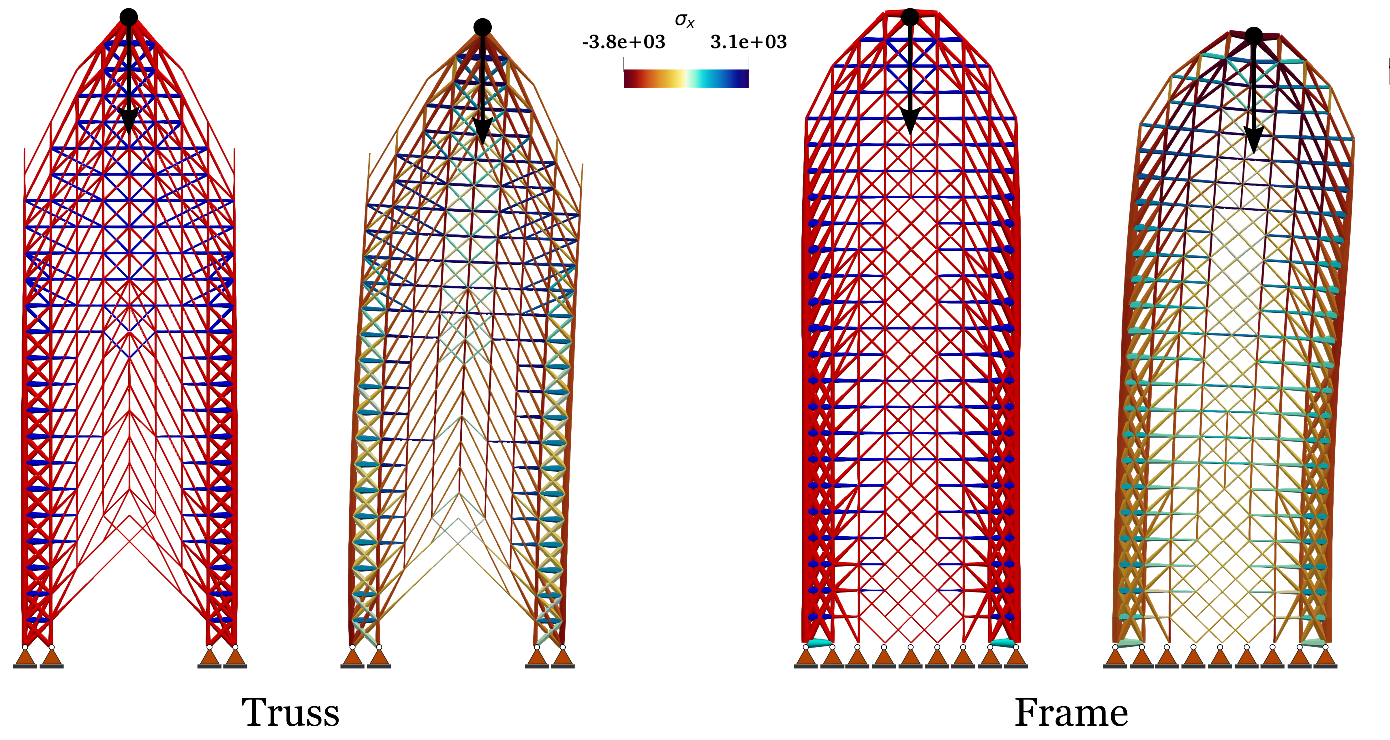}}
 \quad
 \subfloat[]{
  \includegraphics[scale = 0.375, keepaspectratio]
  {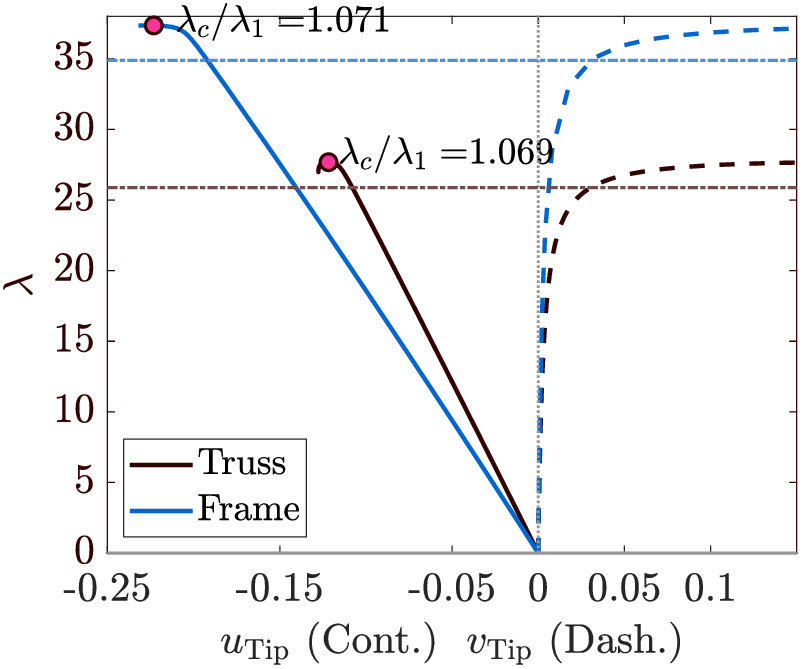}}
 \caption{(a) Column designs obtained when introducing the constraints on the post-buckling slope and curvature in the optimization problem \autoref{eq:optProblemYesStabilityConstraints}. The objective function is linear compliance, the volume fraction $\bar{v}_{f} = 0.25$, and we set $g_{\alpha} = |\alpha|\leq 10^{-10}$, $g_{\beta} = \beta \geq 1.25\beta_{0}$. The buckling constraint is not active for the final designs. (b) shows the nonlinear post-evaluation of the designs' response}
 \label{fig:MotivationExample_Column_designsWithPostBuckling}
\end{figure}

We minimize linear compliance for the volume fraction $\bar{v}_{f} = 0.25$, and increasing values of the lower bound on the BLF ($\lambda_{\rm ref}$). Thus, we solve \eqref{eq:optProblemYesStabilityConstraints} with $W_{0} = \mathbf{f}_{0}^{T}\mathbf{u}_{0}$, and $\lambda_{\rm ref} = \ell\lambda_{1(0)}$, where $\lambda_{1(0)}$ is the BLF of the initial design (see \autoref{tab:Table1}). The optimized truss and frame structures are displayed in \autoref{fig:MotivationExample_Column_optDesigns1}, together with their nonlinear equilibrium paths.

When disregarding the buckling constraint ($\ell = 0$), the designs consist of long compressed columns, narrowly placed near the domain's centerline and connected by only a few tensile bars near the loaded tip. While the compliance is reduced by about $80\%$ and $75\%$, respectively for the truss and frame structures, these have a much lower BLF compared to the initial designs ($\lambda_{1}/\lambda_{1(0)} = 0.156$ and $0.05$). Therefore, we clearly need to introduce the buckling constraint, by setting $\ell > 0$ in \eqref{eq:optProblemYesStabilityConstraints}, and we consider $q = 4$ buckling modes. This drives the optimizer to create wider designs, with more connecting bars between the vertical columns which in turn increases stability against lateral displacements \cite{ferrari-etal_21a, ferrari-sigmund_23a_regularizationLocalizedBuckling}. 

We notice some differences in the designs corresponding to the two types of modeling (see \autoref{fig:MotivationExample_Column_optDesigns1} (a,c)). First, the trusses systematically show more active bars (i.e., bars where $x_{e} > 0$), compared to the frames. In particular, the trusses have $140$, $328$, and $568$ active bars, whereas the frames have $136$, $276$ and $398$ active bars, respectively for $\ell = \{0, 1, 2\}$. Then, the frames show a stronger tapering of the vertical compressed columns, as the beams can better exploit their bending stiffness. On the other hand, the trusses contain more short, $45^{\circ}$ bars, connecting the main struts, especially close to the center of the design domain.

\begin{table}[]
 \centering
  \begin{tabular}{@{}c|cccc|cccc@{}}
   \toprule
   Model          & \multicolumn{4}{c}{Truss} 
                  & \multicolumn{4}{c}{Frame} \\ \midrule
                  & $\ell = 0$ & $\ell = 1.0$ & $\ell = 2.0$ & $g_{\alpha}$ and $g_{\beta}$ &
                    $\ell = 0$ & $\ell = 1.0$ & $\ell = 2.0$ & $g_{\alpha}$ and $g_{\beta}$ \\ \midrule
   $c_{\ast}$     & 31.486     & 32.103       & 33.416       & 37.93                     & 
                    30.311     & 30.788       & 31.699       & 41.65                     \\
   $\lambda_{1}$  &  1.075     &  6.942       & 13.870       & 25.55                     &
                     0.434     &  8.863       & 17.725       & 34.89                     \\
   $\lambda_{c}/\lambda_{1}$
                  &  1.076     &  0.896       & 0.882        & 1.076                     &
                     1.054     &  0.988       & 0.979        & 1.15                      \\
   $|\alpha|$     &  $10^{-10}$&  $3.15\cdot 10^{-4}$  & $6.7\cdot 10^{-2}$ & $10^{-10}$ &
                     $10^{-10}$&  $1.78\cdot 10^{-4}$  & $1.5\cdot 10^{-2}$ & $10^{-10}$ \\
   $\beta$        &  0.0026    &   0.0501     & 0.0129 & 0.0035                          &
                     0.0026    &  -0.0234     & 0.0076 & 0.005                          \\
   \bottomrule
  \end{tabular}
 \caption{Performance of the compressed column designs, obtained for linear compliance minimization with $\bar{v}_{f} = 0.25$, and for increasing values of $\ell$. For the initial design, the compliance, BLF and post-buckling slope and curvature coefficients are $c_{0} = 150.846$, $\lambda_{1(0)} = 6.942$, $\alpha_{0} = 0$, $\beta_{0} = 0.039$, for truss modeling, and $c_{0} = 111.015$, $\lambda_{1(0)} = 8.863$, $\alpha_{0} = 0$, $\beta_{0} = 0.029$, for frame modeling}
 \label{tab:Table1}
\end{table}

Even if these designs meet the minimum imposed BLF, they are very sensitive to imperfections, as can be seen from the nonlinear responses computed in a post-evaluation step, while introducing the transversal perturbation force $\Delta F$. The equilibrium paths in \autoref{fig:MotivationExample_Column_optDesigns1} have been obtained by running the arc-length algorithm up to the load level $\lambda_{\rm max} = 1.05\lambda_{1}$, or until reaching the maximum transversal displacement $|v_{\rm Tip}| = 0.075$. While the initial design reaches the end load $\lambda_{\rm max}$, thus showing stability after the predicted BLF, the designs optimized with buckling constraints fail at a load level $\lambda_{c} < \lambda_{1}$. This occurs systematically for the truss model, where the ratio $\lambda_{c}/\lambda_{1}$ consistently decreases for designs corresponding to higher $\ell$ values (see \autoref{tab:Table1}). The deformation of the truss corresponding to $\ell = 2$ shows a localized failure at the foot, causing a sudden drop of the load carrying capacity, as the bars misalignment triggers a mechanism (see magenta line in \autoref{fig:MotivationExample_Column_optDesigns1}). The situation is less critical for the frame model, as the beams now retain their bending stiffness also at large deformations so we do not observe the same catastrophic collapse. However, the load-displacement curve still has a negative inflection, and therefore the initial post-buckling response is unstable (see $\alpha$ and $\beta$ values in \autoref{tab:Table1}).

We now introduce the constraints \eqref{eq:constraintAlphaBeta} in the optimization formulation \eqref{eq:optProblemYesStabilityConstraints}, thus directly controlling the post-buckling slope and curvature. In particular, we set $\overline{\alpha} = 10^{-10}$ and the minimum value $\underline{\beta} = 1.25\beta_{0}$ for the curvature. The constraint $g_{\alpha}$ serves to indirectly penalize asymmetric responses, which likely appear when considering buckling in the optimization. Indeed, the previous discussion revealed that, if we do not explicitly impose symmetry, it is likely to end up in a slightly non-symmetrical design (thus, $|\alpha| > 0$), even just because of numerical effects.

The designs obtained when considering the post-buckling constraints are displayed in \autoref{fig:MotivationExample_Column_designsWithPostBuckling} (a), and show important differences compared to those in \autoref{fig:MotivationExample_Column_optDesigns1}. First, there are more active bars: $674$ for the truss and $665$ for the frame, and these are more evenly distributed over the design domain. We also notice a different thickness distribution: rather of having a bulk of thick vertical bars localized near the centerline, and many thin connectors, as in the designs of \autoref{fig:MotivationExample_Column_optDesigns1}, now we see a transition of thick bars from the centerline towards the sides of the domain, as we move from the loaded end to the column's foot. Also, there are more tensile bars, and regions with both bars at $\pm 45^{\circ}$ inclination, clearly enhancing the designs' stability against lateral displacements. 

The buckling constraint, which was set at $\ell = 1$, is not active, and these designs indirectly achieve a larger BLF due to the post-buckling constraints (see \autoref{tab:Table1}). This can be explained by recalling that both post-buckling coefficients are defined based on the the BLF and buckling mode, and thus their optimization indirectly affects buckling strength. However, this comes at the price of an increase in the pre-buckling compliance, which, for the truss and frame structures shown in \autoref{fig:MotivationExample_Column_designsWithPostBuckling} is of about $20.5\%$ and $37.4\%$, respectively. Therefore, there is a clear trade-off to consider between the pre- and post-buckling response.

\begin{figure}[t]
 \centering
  \subfloat[]{
  \includegraphics[scale = 0.375, keepaspectratio]
  {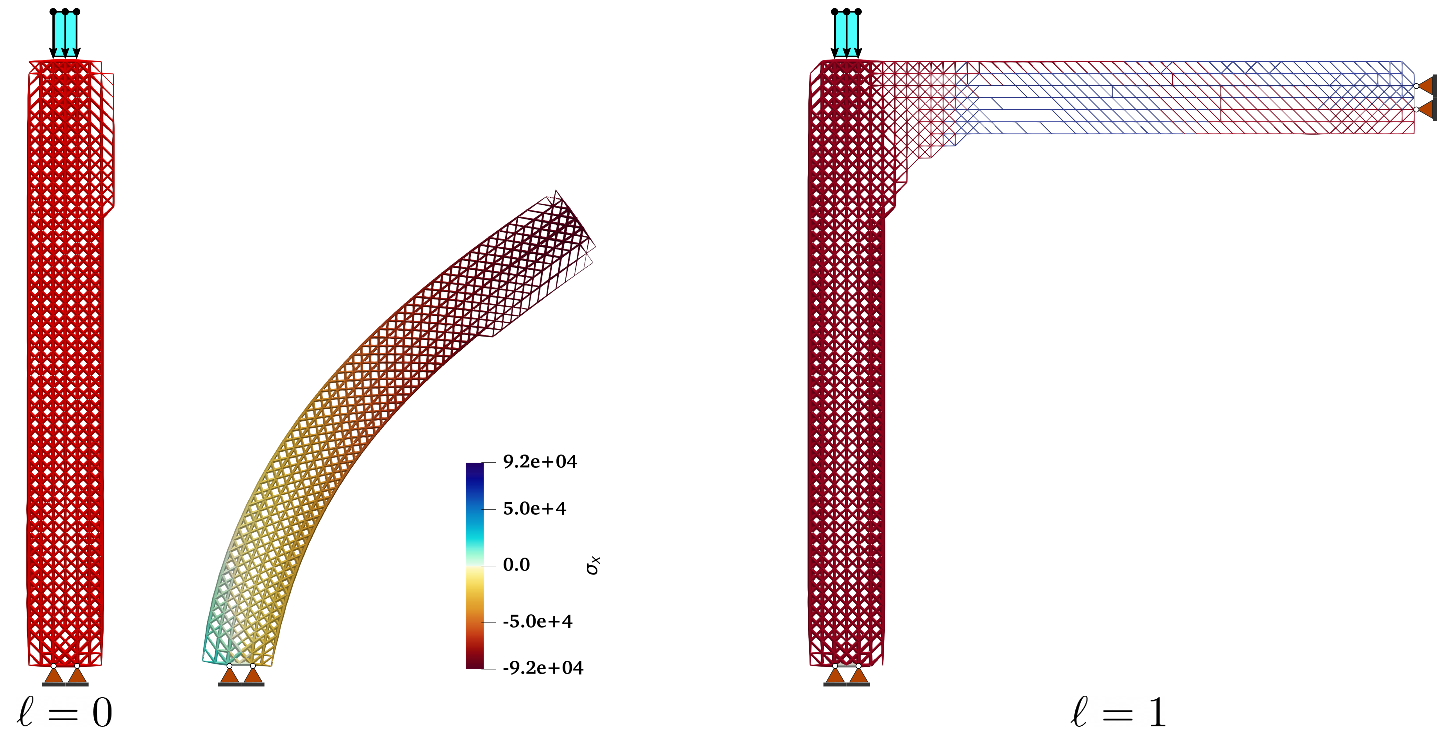}}
  \subfloat[]{
  \includegraphics[scale = 0.40, keepaspectratio]
  {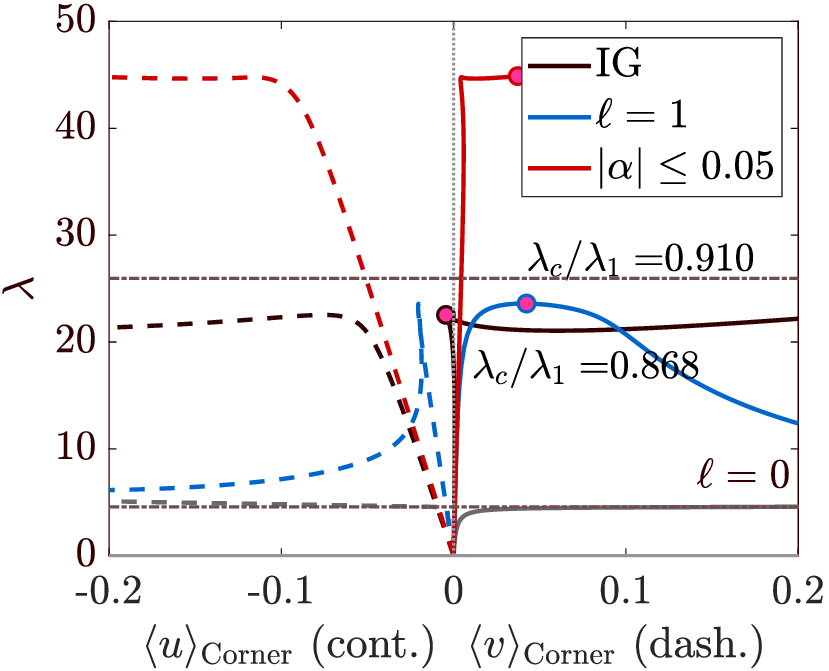}} \\
  \subfloat[]{
  \includegraphics[scale = 0.375, keepaspectratio]
  {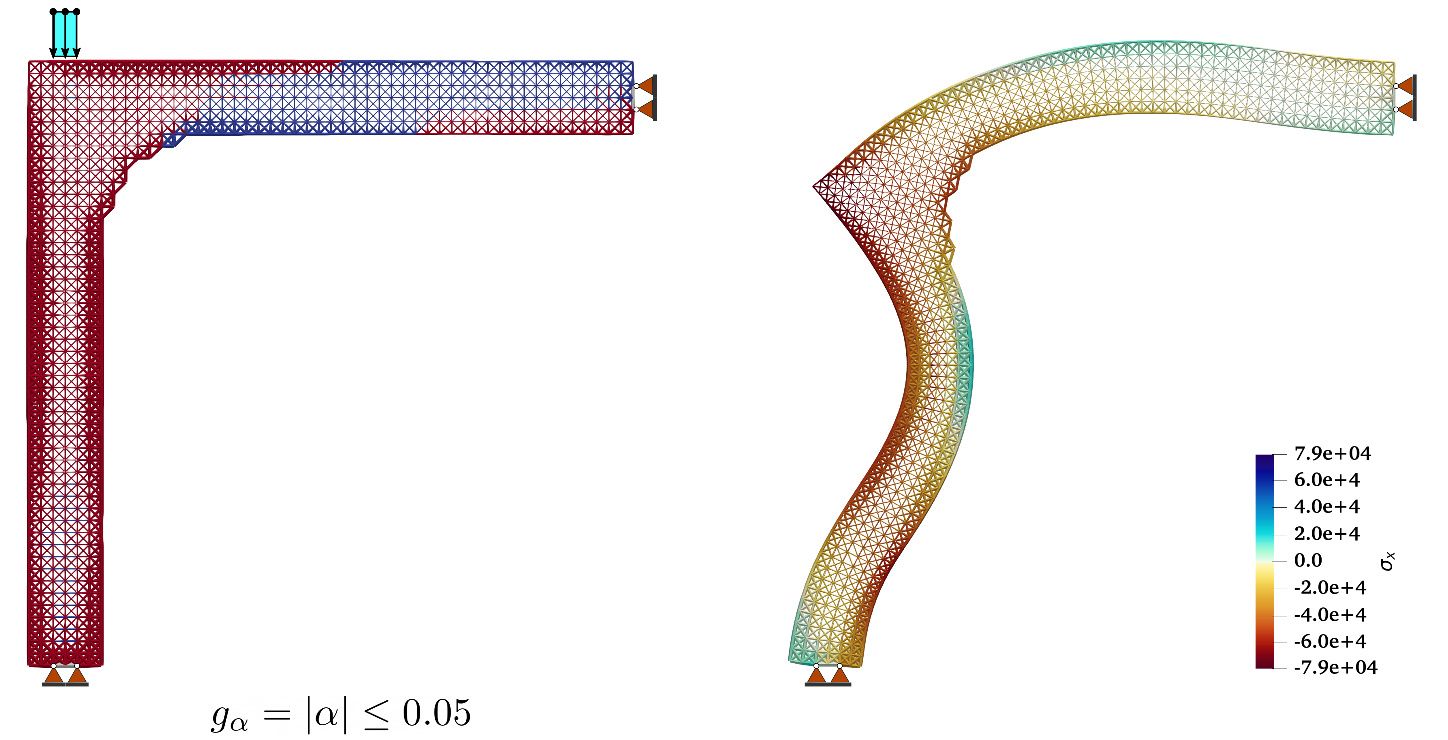}}
  \subfloat[]{
  \includegraphics[scale = 0.40, keepaspectratio]
  {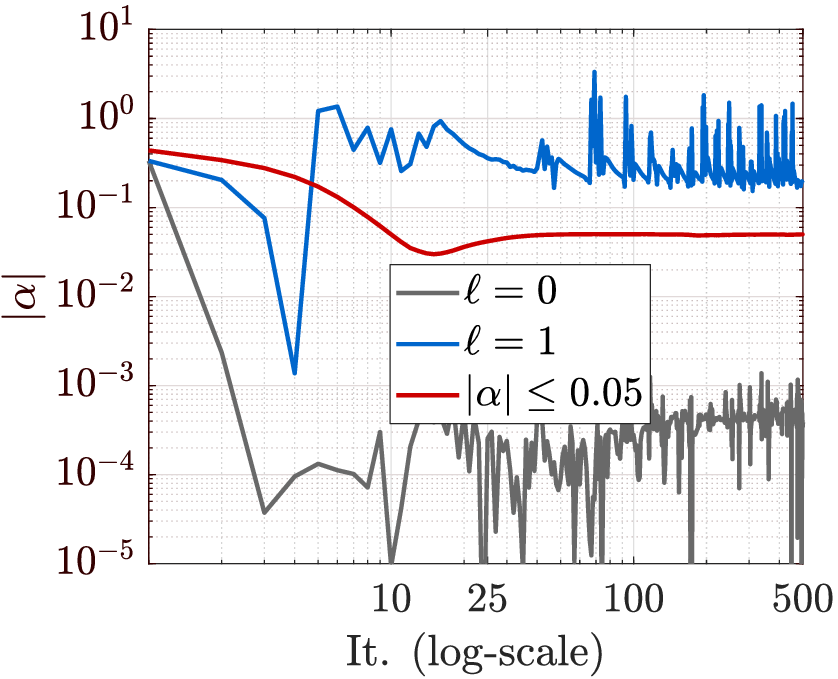}}
 \caption{Results for the Rooda frame example. (a) Basic compliance design (left), and design obtained enforcing the minimum BLF $\lambda_{\rm ref} = \lambda_{1(0)}$ (right) (c) Design obtained by constraining the post-buckling slope to $|\alpha| \leq 0.05$. The deformation plot shows the mean displacement of the loaded nodes $\langle v \rangle_{\rm Corner} = -0.2$, which is also the end value in the nonlinear post-evaluation. (b) post-evaluation of the nonlinear response of the initial and optimized designs. (d) evolution of the post-buckling slope in the optimization proceedings. We highlight that the compliance design would fulfill the condition $|\alpha|<0.05$; however, it clearly has very poor buckling strength}
  \label{fig:Example_RoodaFrame}
\end{figure}

\subsection{Post-buckling slope minimization for the Rooda frame}
 \label{sSec:RoodaFrame}

In order to consider a structure whose imperfection sensitivity is primarily governed by the post-buckling slope $\alpha$, we refer to the Rooda frame in \autoref{fig:NumericalExamplesInitialGuesses} (b), with side lengths $L_{0} = 1$. The assembly consists of $m = 2,425$ beams, each one featuring the material and cross section properties listed in \autoref{sSec:AsymptoticPBanalysisSimpleCases}.

To allow for more design freedom, the width of the two main struts is now set to $t = 0.12$, and both contain 6 square cells of bars with crossing members. The three middle nodes at the end sections of both members are constrained against translations, and a downward load with total magnitude $|F| = 10^{3}$ N is distributed over the three nodes aligned with the supports, at the top corner. An additional load with total magnitude $|\Delta F| = 10^{-3}|F|$, generating a clock-wise couple, is applied when post-evaluating the nonlinear response, not when performing the optimization. We solve \eqref{eq:optProblemYesStabilityConstraints}, minimizing linear compliance $W_{0} = \mathbf{u}^{T}_{0}\mathbf{f}_{0}$, for the volume fraction $\bar{v}_{f} = 0.25$.

The compliance, BLF and post-buckling slope and curvature for the initial design are $c_{0} = 6.218$, $\lambda_{1(0)} = 25.941$, $\alpha_{0} = 0.3343$, and $\beta_{0} = -0.02$. Therefore, the initial design has both non-zero post-buckling slope and negative curvature, and from the nonlinear response, shown as a black curve in \autoref{fig:Example_RoodaFrame} (b), we see that the critical load $\lambda_{c}$ is about $13\%$ lower than the linearly predicted BLF $\lambda_{1}$. Then, even if there is a stiffening behaviour at large displacements, the immediate post-buckling response is unstable.

The compliance design, with no buckling constraint ($\ell = 0$) resembles a straight column, even if the optimizer cannot get completely rid of some bars near the upper right junction. However, its large deformation is clearly that of a column with no lateral restrains (see \autoref{fig:Example_RoodaFrame} (a)). Compared to the initial, uniform design, the compliance is reduced by about $50\%$, but also the BLF is decreased by about $83\%$ ($\lambda_{1} \approx 4.57$). When introducing the buckling constraint $g_{\lambda} = \lambda_{1} \geq \lambda_{1(0)}$ the optimizer clearly tries to stabilize the column against lateral deflections by reintroducing bars in the horizontal part. We obtain the configuration shown on the right of \autoref{fig:Example_RoodaFrame} (a), meeting the buckling constraint and having a compliance slightly higher ($\approx 1.7\%$) compared to the previous one. Yet, the post-buckling coefficients are $\alpha = 0.3548$ and $\beta = -0.04$: even larger than the initial ones, and the post-evaluated critical load, accounting for the perturbation, is sensibly lower ($\approx 10\%$) than the BLF predicted by the eigenvalue buckling analysis (see blue curve in \autoref{fig:Example_RoodaFrame} (b)).

We now introduce the constraint $g_{\alpha} = |\alpha| \leq 0.05$ in the optimization formulation, thus forcing the system to reduce the magnitude of its post-buckling slope. The resulting design is shown in \autoref{fig:Example_RoodaFrame} (c), and we recognize the same trend to build a diffuse reinforcement already encountered in the column example. All the bars are now active ($x_{e} > 0$), and their relative thickness distribution resembles that of a coated structure \cite{clausen-etal_16a}, with a layer of thick bars at the outer edges, on the whole vertical member and close to the corner in the horizontal member. Then, a grid of thin inner bars provides stability to the design, which reaches a very high BLF ($\lambda_{1} = 48.05$), at the price of increasing the compliance by about $60\%$ compared to the value achieved for $\ell = 0$. The post-buckling slope constraint is active ($\alpha = 0.05$), and even if the post-buckling response is not perfectly flat, this design is more stable than all the others, and show a slightly positive post-buckling curvature $\beta = 0.0057$ (see magenta curve in \autoref{fig:Example_RoodaFrame}).

\autoref{fig:Example_RoodaFrame} (d) shows the evolution of $|\alpha|$ in the three optimization runs. Pure compliance optimization leads to a very small value of $|\alpha|$, as the final design is almost symmetric. However, the behaviour is clearly erratic, due to the switching of $\alpha$ between positive and negative values when single bars are removed. Also, the final design shows very poor buckling strength. Optimizing with the buckling constraint also leads to an erratic evolution of the post-buckling slope, which ends up having a magnitude slightly higher than that of the initial design. Finally, the explicit introduction of the constraint $g_{\alpha}$ leads to a smooth reduction of the magnitude of the post-buckling slope, which meets the prescribed upper bound after about 420 optimization steps.

\begin{figure}[t]
 \centering
  \subfloat[$\bar{\lambda} = 0.1\lambda_{c(0)}$]{
  \includegraphics[scale = 0.3, keepaspectratio]
  {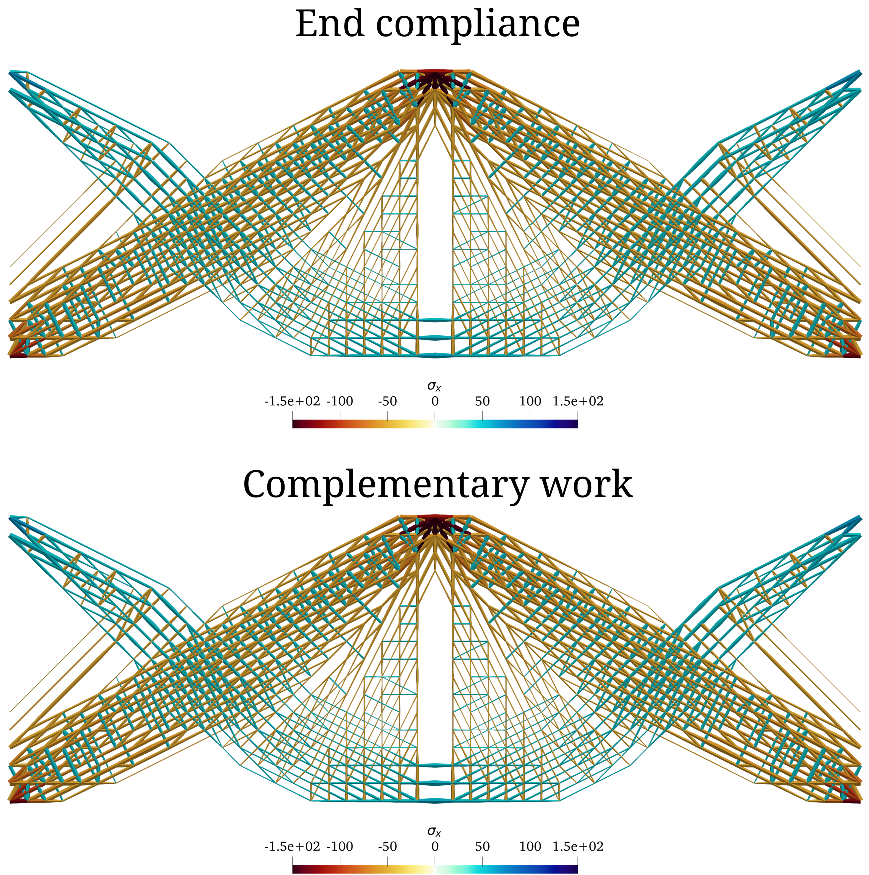}} \qquad
  \subfloat[$\bar{\lambda} = 0.5\lambda_{c(0)}$]{
  \includegraphics[scale = 0.3, keepaspectratio]
  {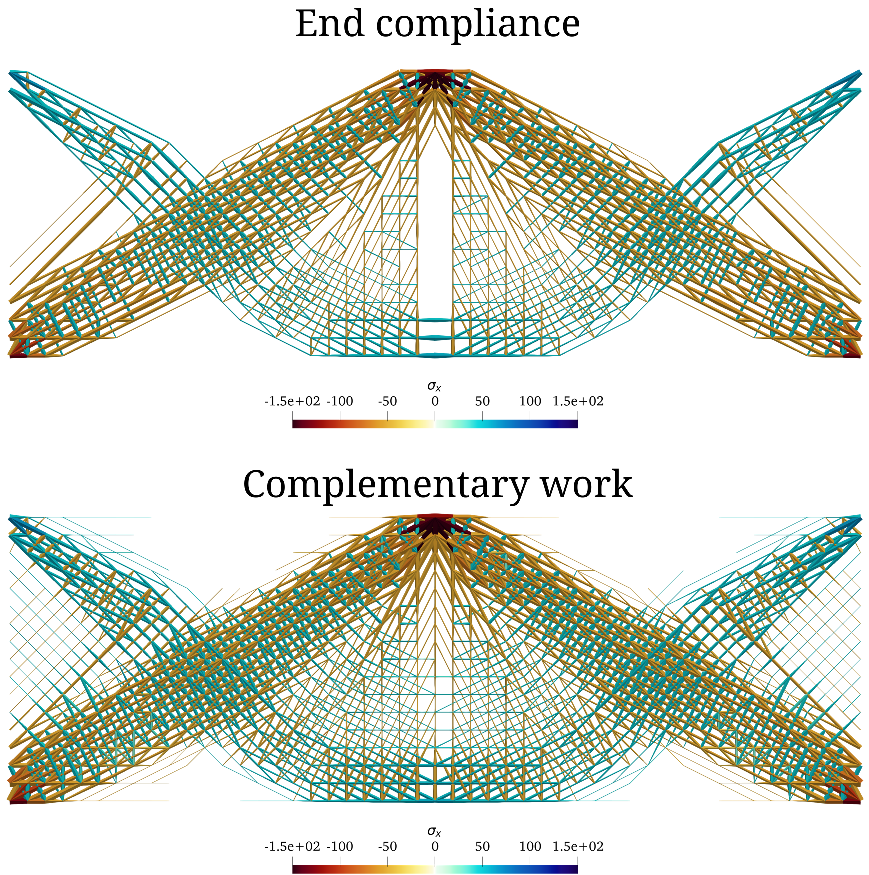}} \qquad
  \subfloat[$\bar{\lambda} = 0.9\lambda_{c(0)}$]{
  \includegraphics[scale = 0.3, keepaspectratio]
  {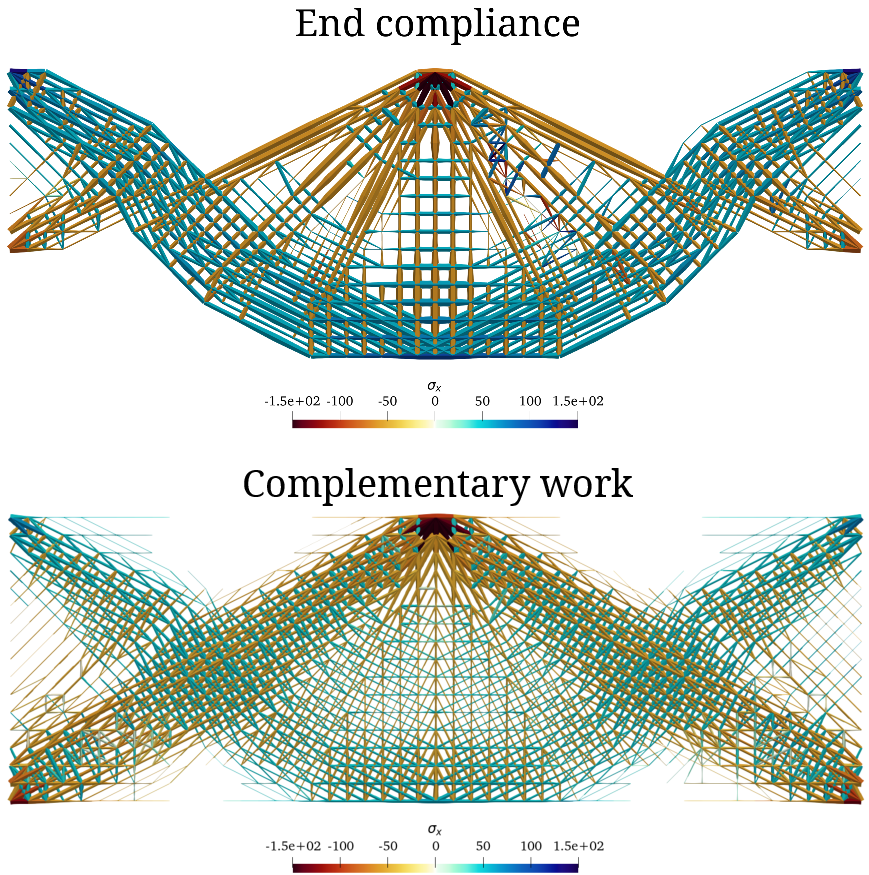}} \\
  \subfloat[]{
  \includegraphics[scale = 0.335, keepaspectratio]
  {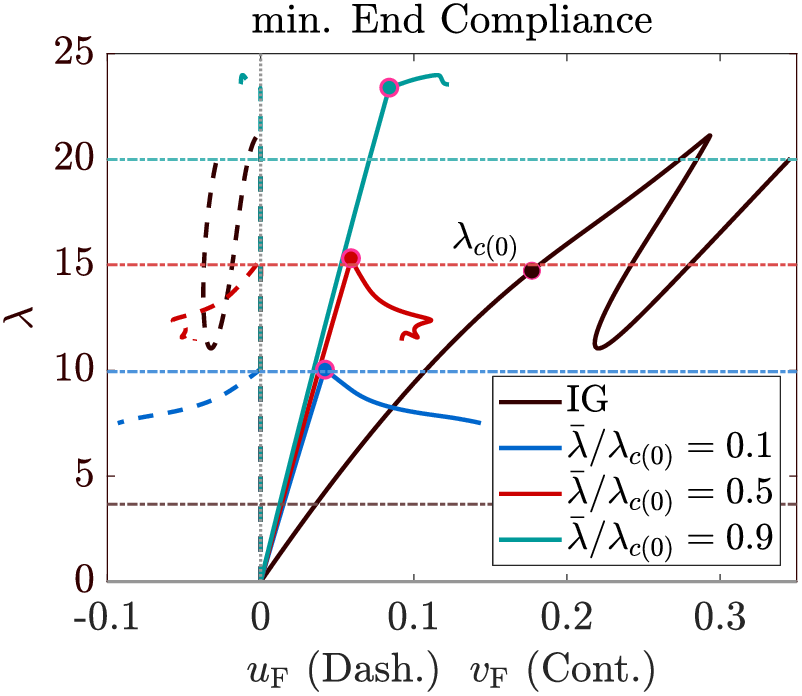}} \quad
  \subfloat[]{
  \includegraphics[scale = 0.335, keepaspectratio]
  {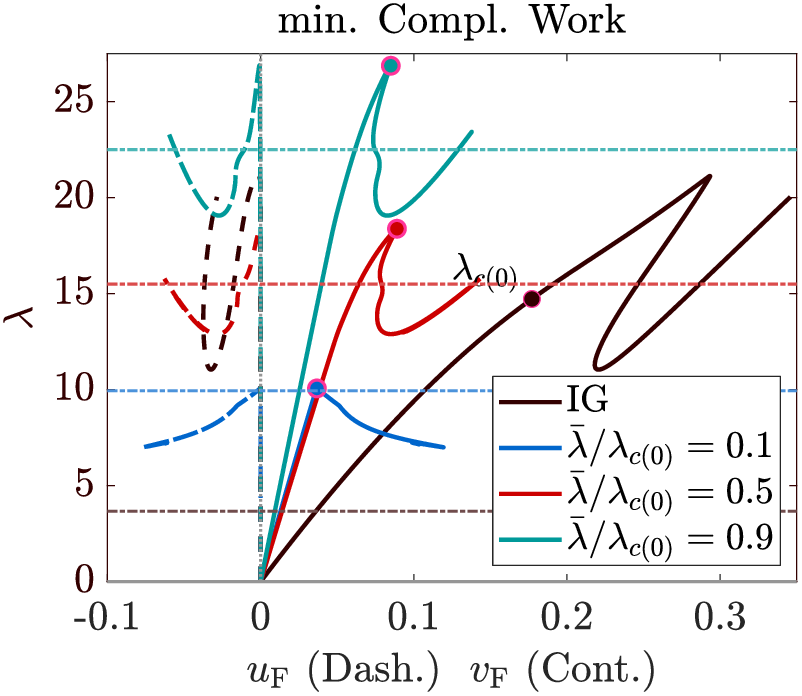}} \quad
  \subfloat[]{
  \includegraphics[scale = 0.305, keepaspectratio]
  {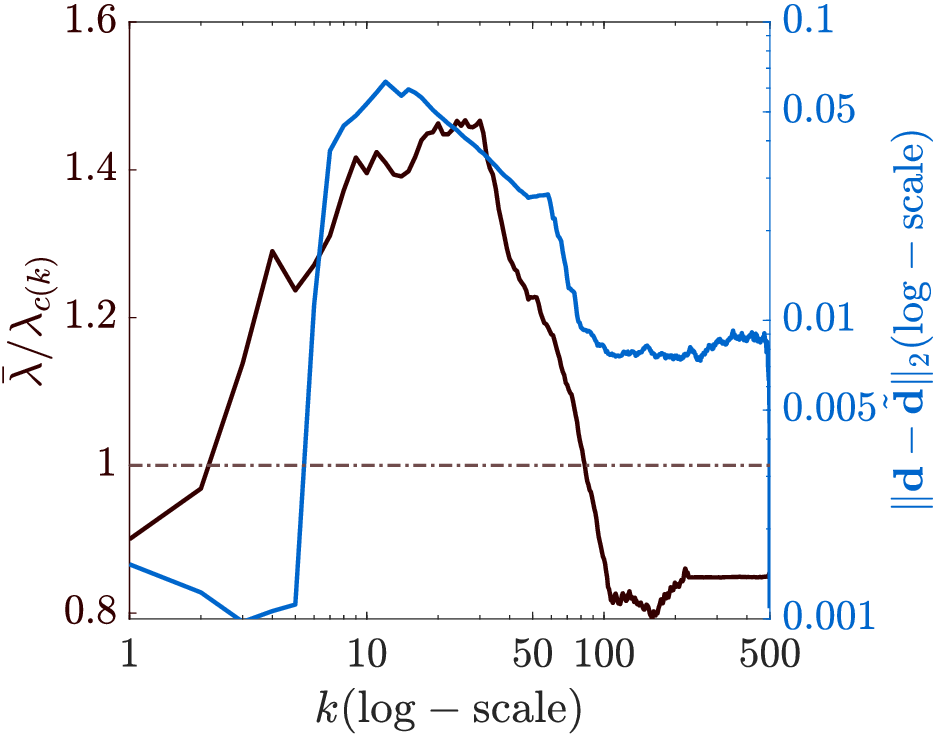}}
  \caption{(a-c) Deep beam designs obtained by minimizing the end-compliance (top) or the complementary energy (bottom). The volume fraction is set to $\bar{v}_{f} = 0.1$, and no buckling constraint is imposed. The nonlinear displacement $\mathbf{d}(\bar{\lambda})$ is computed by using the ANM method. (d) Displays the post-evaluated nonlinear response of the designs optimized for end-compliance. (e) Shows the ratio between the applied and critical loads, and the error norm between the actual displacement and the ANM approximation, at each iteration, when minimizing $W_{EC}$ at $\bar{\lambda}/\lambda_{c(0)} = 0.9$}
 \label{fig:Example_DeepBeam}
\end{figure}

\subsection{Optimization of the nonlinear response of a double clamped deep beam}
 \label{sSec:ClampedDeepBeam}

We consider the double clamped beam sketched in \autoref{fig:NumericalExamplesInitialGuesses} (c), which is inspired by the configuration studied in \cite{buhl-etal_00a_geometricNonlinearTO}. The domain has aspect ratio 3:1, and the nodes define a grid of $48\times 16$ squares with side length $h$, and a total of $m = 6,080$ bars connect nodes within a radius of $r = 2\sqrt{2}h$. The assembly is modeled as a frame, and the $(u,v)$ displacements are restrained on all the nodes at the left and right edges. A downward force with magnitude $|F| = 10^{3}$ N is applied to the mid node at the top edge. Then, an additional downward triangular force distribution, with total magnitude $|\Delta F| = 10$ N and generating a clockwise couple, is applied to the loaded node and to the two adjacent ones. For this example, two load cases with total force $F \pm \Delta F$ will be considered in the optimization, since we are directly optimizing the nonlinear response.

The black curves in \autoref{fig:Example_DeepBeam} (d) show the nonlinear equilibrium path for the initial design. As for the previous examples, the horizontal lines mark the BLFs predicted by the linearized eigenvalue analysis \eqref{eq:APP_discretized_PerturbationEquationsMultimodal}$_{2}$, whereas the dots mark the first load level at which $|K_{t}| = 0$. For the initial design we have $\lambda_{c}/\lambda_{1} \approx 4.1$, so the linear BLF is far lower than the actual critical load ($\lambda_{c(0)}$).

This shows the limitations of the linearized pre-buckling stress assumptions in representing the nonlinear stiffening effect at large bending deformations, and its impact on the local failure of a few bars near the loaded point, which in reality is mainly governed by the material nonlinearity \cite{pedersen_03a_toFramePathDependent}. Thus, for this example we need to compute the critical pair $(\lambda_{1},\mathbf{v}_{1})$ from the nonlinear eigenvalue problem $K_{t}(\mathbf{u}_{0}(\lambda_{1}))\mathbf{v}_{1} = 0$, following from the discretization of \autoref{eq:firstOrder_eigenvalueEquation}.

We aim at minimizing the end-compliance $W_{EC}$, or maximizing the complementary work $W_{CE}$, at a given load level $\bar{\lambda}$ (see \autoref{eq:nonLinearResponses_ECCW}), for the volume fraction $\bar{v}_{f} = 0.1$. To this end we use the ANM described in \autoref{Sec:FormulationAsymptoticPostBuckling}, here considering a single buckling mode, to approximate the actual nonlinear displacement, say $\tilde{\mathbf{d}}(\xi)$, where the coordinate $\xi$ is given the relationship $\bar{\lambda} = \bar{\lambda}(\xi)$.

\autoref{fig:Example_DeepBeam} shows the designs corresponding to three load levels $\bar{\lambda} = \{0.1, 0.5, 0.9\}\lambda_{c(0)}$, where $\lambda_{c(0)}$ is the critical load of the initial design, computed by the nonlinear eigenvalue analysis. The design in (a) is very close to the linear compliance minimization one, which resembles a Mises truss, although we see several bars in tension. These are generally missing in continuum-based solutions, which consist of two mainly compressed struts \cite{buhl-etal_00a_geometricNonlinearTO, dalklint-etal_20_removingVoiElements, dunning_23a_stabilityConsraintsGeometricNonlinearTO}. However, the frame modeling makes it harder to obtain a purely compressed design, thus keeping some tensile bars. The critical load of this design is lower than the initial one, and is now very well predicted by the linearized BLF.

As the level of the final load increases, we observe the development of more and more bars in tension, and the design corresponding to $\bar{\lambda} = 0.9\lambda_{c(0)}$ mainly consists of an inverted arch, in line with what shown by \cite{buhl-etal_00a_geometricNonlinearTO, kemmler-etal_05a_largeDeformationStability}. Also, increasing $\bar{\lambda}$ is followed by an increase of the critical load of the optimized design, and when $\bar{\lambda} = 0.9\lambda_{c(0)}$ we also achieve some post-buckling stability.

The designs obtained minimizing the complementary work are all very similar to each other, and we see the tendency to introduce more bars in the design domain. These achieve a larger critical load, compared to the end compliance ones. However, without direct control of the post-buckling coefficient, they show a sudden post-buckling failure.

\autoref{fig:NumericalExamplesInitialGuesses} (e) shows the ratio of the applied load level, with respect to the current critical load $\bar{\lambda}/\lambda_{c}$ for each step of the optimization, and for the case $\bar{\lambda} = 0.9\lambda_{c(0)}$. We see that this value occasionally become larger than one, especially in the initial iterations, meaning that the displacement asymptotic approximation happens in the actual post-buckling regime. Nevertheless, the accuracy of such approximation is always very good, as the quantity $\|\mathbf{d}-\tilde{\mathbf{d}}\|_{2}$ is always below a few percent (see blue curve in \autoref{fig:NumericalExamplesInitialGuesses} (e)).

\section{Discussion and outlook}
 \label{Sec:DiscussionOutlook}

We have discussed the use of an asymptotic-numerical method (ANM), based on the post-buckling perturbative theory, for sizing and topology optimization. First, we have shown the need to account for the post-buckling response in the optimization problem, to achieve designs which are robust against imperfections and avoid catastrophic failure. We remark that the introduction of a simple buckling constraints does not necessarily lead to robust designs, especially for configurations with little inner redundancies, such as lattices with low volume fractions. In such cases, buckling-driven optimization might even worsen the structures' imperfection sensitivity.

The asymptotic theory offers a computationally efficient, yet rigorous tool for controlling the systems' initial post-buckling response. A recent approach, based on minimizing the compliance for a weighted sum of the pre-buckling and buckling displacements \cite{victoria-etal_23a_IncorporatingBucklingTOisolines} shares some similarities with the present method. However, the clear advantages of the perturbation approach are: (1) it naturally provides the correct physical weighting between the pre- and post-buckling displacements; and (2) it allows enforcing constraints on the asymptotic coefficients governing the initial post-buckling stability and imperfection sensitivity.

\paragraph{Computational considerations} The ANM provides an approximation to the structural nonlinear response, both in the pre- and post-buckling regimes, which otherwise has to be computed by potentially expensive and cumbersome nonlinear analyses.

Some details on the FE implementation of \eqref{eq:zeroOrder_equilibrium}-\eqref{eq:asymptoticCoefficients} are given in \ref{App:DetailsFEM_1}. Compared to a classical eigenvalue buckling analysis, which already requires solving \eqref{eq:zeroOrder_equilibrium} and \eqref{eq:firstOrder_eigenvalueEquation}, the setup of the ANM only involves limited additional efforts, linked to the solution of the extra system for the second-order displacement correction, \eqref{eq:secondOrderDisplacementCorrection}. Then, computing the post-buckling slope and curvature only requires a few matrix-vector products.

The sensitivity analysis of the post-buckling coefficients involves the derivatives of the eigenvectors and of the second order displacement fields, and this adds some complexity compared to the case of simple buckling optimization. However, efficient methods based on eigenvector aggregates \cite{tcherniak_02a_resonatingStructuresTOsimp, li-etal_23a_eigenvectorAggregateTO} can be used for this task.

The computational cost clearly increases when facing large nonlinearities already in the pre-buckling regime. In such a case, the clamped beam example of \autoref{sSec:ClampedDeepBeam} shows that the linearized buckling analysis would yield a wrong prediction of the buckling mode $\mathbf{v}_{1}$, consequently impairing the accuracy of the asymptotic expressions. Therefore, the linearized eigenvalue problem \eqref{eq:APP_discretized_PerturbationEquationsMultimodal}$_{2}$ should be replaced by the nonlinear one $K_{t}(\mathbf{u}_{0}(\lambda))\mathbf{v}_{1} = 0$, where $K_{t}$ is the full tangent stiffness matrix, depending at least quadratically on $\lambda$. However, the same complication is shared also by ``classical'' approaches to the solution of the nonlinear and post-buckling problem, based on the incremental-iterative solution of \eqref{eq:stateEquation_nonlinear}-\eqref{eq:newtonStep}, and on tracking of the critical points either by direct \cite{wriggers-etal_88a_quadraticallyConvergentProcedureStability, kemmler-etal_05a_largeDeformationStability} or approximate methods \cite{dalklint-etal_20a_eigenfrequencyHyperelastic, dunning_23a_stabilityConsraintsGeometricNonlinearTO}. In particular, the robust and efficient determination of critical points is still a challenging point, faced by all classical nonlinear solvers \cite{dunning_23a_stabilityConsraintsGeometricNonlinearTO}.

Overall, in our numerical tests the ANM turned out to be  10-15 times faster compared to the classical arc-length solution. For example, in the complementary work minimization for the deep beam example, where the whole nonlinear equilibrium path is needed, each redesign step takes, on average, $0.55$s when using the ANM, and $8.4$s when using the path-following algorithm.

\paragraph{Current limitations and extension to continuum TO} The real hurdle for the implementation of the ANM is linked to the choice of a suitable structural modeling, avoiding spurious strains and deformation locking. This is a critical point, especially for models based on kinematic constraints, such as beams and plates, where theories based on non-conventional strain measures and discretization techniques may be required to achieve a high accuracy of the post-buckling curvature and equilibrium path. Even if it does not increase the overall computational cost, this requires the development of \emph{ad hoc} procedures, often not included in mainstream FE packages \cite{dellarosa_damkilde_2014_Thesis}.

Also, we recall that the ANM provides a high quality approximation of the real response only in a neighborhood of the bifurcation point \cite{brezzi-etal_86a_getAroundQuadraticFold}. However, this is generally sufficient to investigate the \textit{initial} post-buckling response, whereas the accuracy of the ANM can be improved further by coupling with other expansion methods \cite{cochelin-etal_94a_asymptoticNumericalPade}.

Finally, the use of asymptotic-numerical procedures seems very promising, especially for continuum TO, where the increased modeling efforts are easily paid off by cutting the huge computational cost required by large-scale incremental nonlinear analyses. Also, in the continuum setting we expect spurious deformation to be less of an issue, since there are generally less kinematic restrictions. The need to use mixed discretizations, avoiding locking phenomena may still remain; however, this can easily be achieved by conventional techniques \cite{pian-sumihara_84a, wilson-ibrahimbergovic_90a}.

\appendix
\renewcommand{\thefigure}{A\arabic{figure}}
 \setcounter{figure}{0}

\section{Asymptotic-numerical post-buckling method for multiple buckling modes}
 \label{App:DetailsFEM_2}

We extend the formulations of \autoref{Sec:FormulationAsymptoticPostBuckling} to the case of $q$ buckling modes, either coinciding or not. A more detailed treatment of this case can be found, e.g., in \cite{byskov-hutchinson_77a_modeInterationAxiallyStiffened, olesen-byskov_82a_asymptoticPostbucklingStresses, salerno-lanzo_97a_nonlinearFElockingFree, NOTE_garcea-etal_13a_postBucklingAsymptotic}. To simplify the equations, here we use the notation of \cite{byskov-hutchinson_77a_modeInterationAxiallyStiffened}, where $l_{1}(\mathbf{u})$ and $l_{11}(\mathbf{u},\mathbf{v})$ are the linear and bilinear generalized strain operators associated with displacements $\mathbf{u}$ and $\mathbf{v}$, and $\boldsymbol{\sigma}$ is the generalized stress. The summation convention is implied for all repeated indices but those put between brackets.

The asymptotic expansion \eqref{eq:parametricExpression_DLambda} becomes
\begin{equation}
 \label{eq:APPexpansionMultipleModes}
 \begin{aligned}
  & \mathbf{d}(\xi) = \lambda(\xi)\mathbf{u}_{0} + \xi_{i}
  \mathbf{v}_{i} + \xi_{i}\xi_{j}\mathbf{w}_{ij} \\
  & \xi_{(i)}\left(1-\frac{\lambda}{\lambda_{(i)}}\right) + \xi_{i}\xi_{j}\alpha_{(i)ij} + \xi_{i}\xi_{j}\xi_{k}\beta_{(i)ijk} = 0 \: , \qquad (i) = 1, \ldots, q
 \end{aligned}
\end{equation}
where $\mathbf{u}_{0}$ is the pre-buckling displacement, ($\mathbf{v}_{i}, \lambda_{i}$) are the multiple bifurcation modes and loads, and $\mathbf{w}_{ij}$ the second order displacement corrections corresponding to the $i$-th and $j$-th modes.

The equilibrium path $(\lambda, \boldsymbol{\xi})\in\mathbb{R}\times \mathbb{R}^{q}$ is now governed by $q$ nonlinear equations, featuring the sets of asymptotic coefficients $\alpha_{(i)ij}$ and $\beta_{(i)ijk}$
\begin{equation}
 \label{eq:APPasymptoticCoefficients}
 \begin{aligned}
  \alpha_{(i)ij} & = \boldsymbol{\sigma}_{(i)}\cdot l_{11}(\mathbf{v}_{i},\mathbf{v}_{j}) + 
  2 \boldsymbol{\sigma}_{j}\cdot l_{11}(\mathbf{v}_{i},\mathbf{v}_{(i)}) \\
  \beta_{(i)ijk} & = \boldsymbol{\sigma}_{(i)i}\cdot l_{11}(\mathbf{v}_{j},\mathbf{v}_{k}) + 
  \boldsymbol{\sigma}_{ij}\cdot l_{11}(\mathbf{v}_{k},\mathbf{v}_{(i)}) + \boldsymbol{\sigma}_{(i)}\cdot l_{11}(\mathbf{v}_{i},\mathbf{w}_{jk}) + \boldsymbol{\sigma}_{i}\cdot l_{11}(\mathbf{v}_{(i)},\mathbf{w}_{jk}) + \\
  & 2 \boldsymbol{\sigma}_{i}\cdot l_{11}(\mathbf{v}_{j},\mathbf{w}_{k(i)})
 \end{aligned}
\end{equation}
corresponding to each buckling mode $(i)$, and depending on each pair $(i,j)= 1\ldots, q$, or triple $(i,j,k) = 1, \ldots, q$ of buckling and second order displacements.

In \eqref{eq:APPasymptoticCoefficients} we have assumed the normalization $\boldsymbol{\sigma}_{0}\cdot l_{11}(\mathbf{v}_{(i)},\mathbf{v}_{(i)}) = 1$, and a stress with double subscript is based on the corresponding second order displacement. The perturbation equations \eqref{eq:zeroOrder_equilibrium}-\eqref{eq:secondOrderDisplacementCorrection} now read
\begin{align}
 \label{eq:APPzeroOrderProblemMultimodal}
 \boldsymbol{\sigma}_{0}\cdot l_{1}(\delta\mathbf{d}) - \mathbf{f}_{0}\cdot \delta\mathbf{d} & = 0 
 \: , \qquad \forall\:\delta\mathbf{d} \\
 \label{eq:APPfirstOrderProblemMultimodal}
 \boldsymbol{\sigma}_{i}\cdot l_{1}(\delta\mathbf{d}) + \lambda_{i} \boldsymbol{\sigma}_{0} \cdot l_{11}(\mathbf{v}_{i}, \delta\mathbf{d}) & = 0
 \: , \qquad \forall\:\delta\mathbf{d} \: , \mathbf{v}_{i} \neq 0 \\
 \label{eq:APPsecondOrderProblemMultimodal}
 \boldsymbol{\sigma}_{ij}\cdot l_{1}(\delta\mathbf{d}) + \lambda_{1}\boldsymbol{\sigma}_{0}\cdot l_{11}(\mathbf{w}_{ij},\delta\mathbf{d}) +
 \frac{\boldsymbol{\sigma}_{i}\cdot l_{11}(\mathbf{v}_{j},\delta\mathbf{d}) +
 \boldsymbol{\sigma}_{j}\cdot l_{11}(\mathbf{v}_{i},\delta\mathbf{d})}
 {2} & = 0 \: , \qquad \forall\: \delta\mathbf{d}
\end{align}

Despite the seemingly complicated expressions, the increase in the computational cost, compared to \eqref{eq:zeroOrder_equilibrium}-\eqref{eq:secondOrderDisplacementCorrection}, is only minor. The pre-buckling equilibrium equation is same as before, and so is the eigenvalue equation \eqref{eq:APPfirstOrderProblemMultimodal}, as we need to solve it for several modes anyways. The second order problem \eqref{eq:APPsecondOrderProblemMultimodal} now gives a set of $\frac{q(q+1)}{2}$ equations, due to the symmetry of the right hand sides, which are completed by the orthogonality conditions $\boldsymbol{\sigma}_{0}\cdot l_{11}(\mathbf{w}_{ij}, \mathbf{v}_{k}) = 0$.

\section{Expressions in the algebraic discretized setup}
 \label{App:DetailsFEM_1}
 
Within the FEM setup equations \eqref{eq:APPzeroOrderProblemMultimodal}-\eqref{eq:APPsecondOrderProblemMultimodal} give the following algebraic systems
\begin{equation}
 \label{eq:APP_discretized_PerturbationEquationsMultimodal}
 \begin{aligned}
  K_{0}\mathbf{u}_{0} & = \mathbf{f}_{0} \\
  \left[ K_{0} + \lambda_{i} G(\mathbf{u}_{0})\right]\mathbf{v}_{i} & = 0 \:, \qquad \mathbf{v}_{1} \neq 0 \\
  \left[ K_{0} + \lambda_{1} G(\mathbf{u}_{0})\right]\mathbf{w}_{ij} & = 
  \frac{1}{2}\left[
  G(\mathbf{v}_{i})\mathbf{v}_{j} + G(\mathbf{v}_{j})\mathbf{v}_{i} + \sum^{q}_{k=1}\alpha_{ijk}\lambda_{k}
  G(\mathbf{u}_{0})\mathbf{v}_{k}
  \right]
 \end{aligned}
\end{equation}
where the linear stiffness and geometric matrices are computed as detailed in \ref{App:DetailsFEM}. Then, the post-buckling coefficients are computed by inexpensive matrix-vector products
\begin{equation}
 \label{eq:discretized_AlphaBeta}
 \begin{aligned}
  \alpha_{(i)ij} & = \frac{\mathbf{v}^{T}_{j}G(\mathbf{v}_{(i)})\mathbf{v}_{i} + 
  2 \mathbf{v}^{T}_{(i)}G(\mathbf{v}_{j})\mathbf{v}_{i}}
  {2\lambda_{(i)}} \\
  \beta_{(i)jkl} & = \frac{\mathbf{v}^{T}_{j}G(\mathbf{v}_{(i)i})\mathbf{v}_{k} + 
  \mathbf{v}^{T}_{(i)}G(\mathbf{v}_{ij})\mathbf{v}_{k} + \mathbf{v}^{T}_{i}G(\mathbf{v}_{(i)})\mathbf{w}_{jk} + \mathbf{v}^{T}_{(i)}G(\mathbf{v}_{i})\mathbf{w}_{jk} + 2\mathbf{v}^{T}_{j} G(\mathbf{v}_{i})\mathbf{w}_{(i)k}}
  {2\lambda_{(i)}}
 \end{aligned}
\end{equation}

The above equations are all is needed for the post-buckling analysis of the perfect structure, while considering linearized pre-buckling: one linear analysis, one standard eigenvalue analysis, and an extra linear analysis featuring the same system matrix and multiple right hand sides.

Then, the introduction of imperfections is basically inexpensive, as it only requires the modification of the right hand side of \eqref{eq:APPexpansionMultipleModes}. Assuming that the imperfection shape $\Delta$ can be decomposed over the set of $q$ buckling modes $\Delta = \Delta_{i}\mathbf{v}_{i}$, the right hand side of \eqref{eq:APPexpansionMultipleModes} becomes $\lambda\Delta_{i}\lambda^{-1}_{(i)}$.

The reduction to the case of a single buckling mode simplifies the second order problem as 
\begin{equation}
 \label{eq:discretized_PerturbationEquations}
  \left[ K_{0} + \lambda_{1} G(\mathbf{u}_{0})\right]\mathbf{w} = - G(\alpha\lambda_{1}\mathbf{u}_{0} + \mathbf{v}_{1})\mathbf{v}_{1} - \boldsymbol{\chi}_{1}\mathbf{v}^{T}_{1}G\mathbf{v}_{1}
\end{equation}
and the expressions of the post-buckling coefficients
\begin{equation}
 \label{eq:discretized_AlphaBeta}
 \begin{aligned}
  \alpha & = -\frac{3\boldsymbol{\sigma}_{1}\cdot
  l_{2}(\mathbf{v}_{1})}
  {2\lambda_{1}}
  \simeq -\frac{3}{2\lambda_{1}}\mathbf{v}^{T}_{1}G(\mathbf{v}_{1})\mathbf{v}_{1} \\
  \beta & = -\frac{\boldsymbol{\sigma}_{11}\cdot l_{2}(\mathbf{v}_{1}) + 
  2\boldsymbol{\sigma}_{1}\cdot l_{1}(\mathbf{w},\mathbf{v}_{1})}
  {\lambda_{1}}
  \simeq -\frac{\mathbf{v}^{T}_{1}G(\mathbf{w})\mathbf{v}_{1} + 
  \mathbf{w}^{T}G(\mathbf{v}_{1})\mathbf{v}_{1}}
  {\lambda_{1}\mathbf{v}^{T}_{1}G(\mathbf{u}_{0})\mathbf{v}_{1}}
 \end{aligned}
\end{equation}

\section{Details on the structural models and their discretization}
 \label{App:DetailsFEM}

\begin{figure}[t]
 \centering
  \subfloat[]{
  \includegraphics[scale = 0.4, keepaspectratio]
  {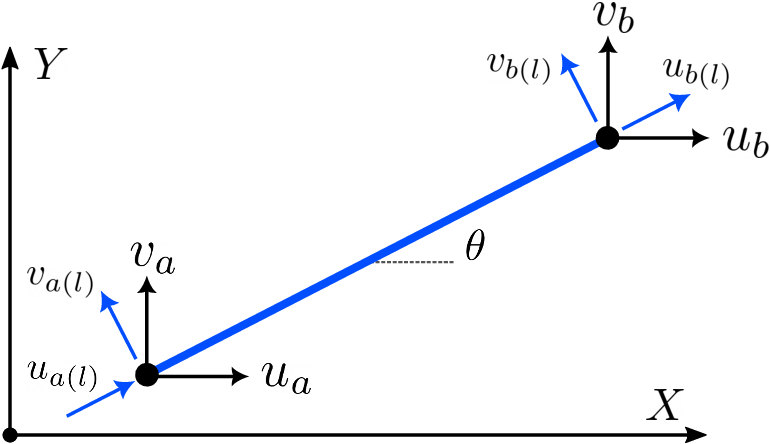}} \qquad
  \subfloat[]{
  \includegraphics[scale = 0.4, keepaspectratio]
  {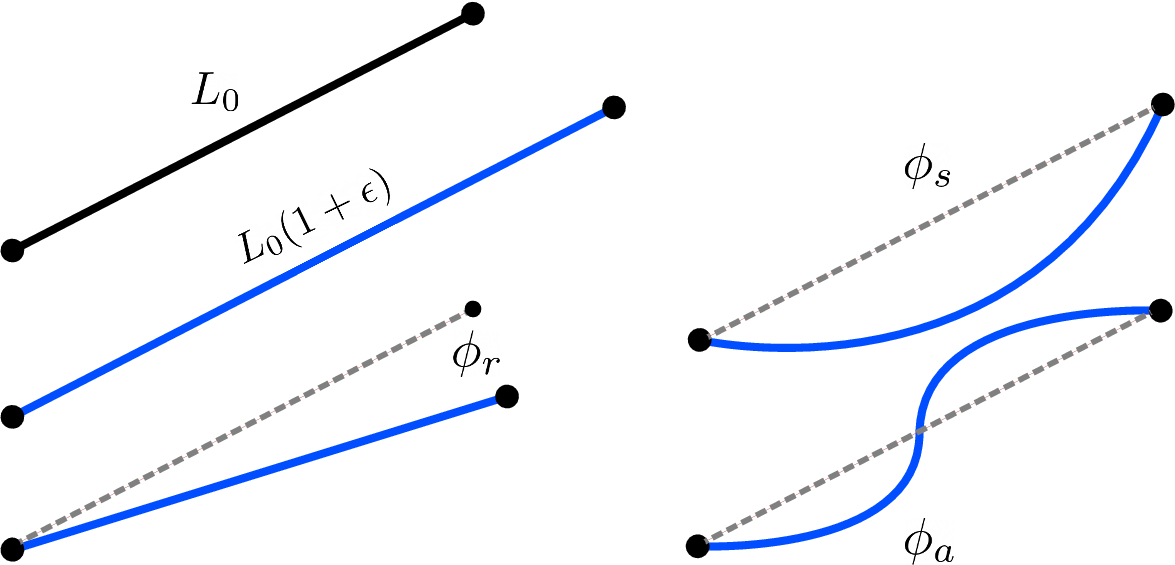}} 
  \caption{Sketch of the bar element with the kinematic DOFs (a), and illustration of the \textit{natural deformation} modes \cite{argyris-etal_79a_naturalModesFEM}. The beam model accounts for all the DOFs and natural deformations displayed, whereas the rod model does not possess rotations, and the associated symmetric and asymmetric bending modes ($\phi_{s}$, $\phi_{a}$)}
 \label{fig:APProdandbeamsketch}
\end{figure}

Here we provide some details about the rod and beam models used in the asymptotic-numerical method. The literature on this topic is vast, and we refer to \cite{book:crisfield91} for the basic conventions an discussion on strain measures for rods and beams. In both cases, it is convenient to formulate the elements in terms of natural deformations \cite{argyris-scharpf_68a_naturalModesTechnique_part1, argyris-scharpf_68a_naturalModesTechnique_part2, argyris-etal_79a_naturalModesFEM} (see \autoref{fig:APProdandbeamsketch}).

\subsection{Rod element used for truss modeling}
 \label{sApp:rodElement&discretization}

Adopting the Green-Lagrange strain measure $\epsilon = \epsilon_{0} + u_{x} + \frac{1}{2}(u^{2}_{x} + v^{2}_{x})$, where $\epsilon_{0}$ is a given pre-strain, the strain energy of the rod reads ($d = (u,v)$)
\begin{equation}
 \label{eq:APPstrainEnergyRod}
 \begin{aligned}
  \Phi(d) & = \frac{1}{2} \int^{L_{0}}_{0} EA
  \left[\epsilon_{0}+u_{x}+\frac{1}{2}(u^{2}_{x} + v^{2}_{x})\right]^{2}
  {\rm d}x \\ & =
  \frac{1}{2} \int^{L_{0}}_{0}EA\left[
  \epsilon^{2}_{0}+
  2\epsilon_{0}u_{x}+
  u^{2}_{x}+\epsilon_{0}(u^{2}_{x}+v^{2}_{x})+
  u_{x}v^{2}_{x}+u^{3}_{x}+
  \frac{1}{2}u^{2}_{x}v^{2}_{x}+\frac{1}{4}(u^{4}_{x}+v^{4}_{x})
  \right] {\rm d}x
  \end{aligned}
\end{equation}

Introducing the test fields $\delta d$, $\tilde{d}$, $\hat{d}$ and $\bar{d}$, the first four variations of \eqref{eq:APPstrainEnergyRod} read
\begin{equation}
 \label{eq:APPstrainEnergyRod_variations}
 \begin{aligned}
  \Phi^{(1)}_{0}[\delta d]
  & = \int^{L_{0}}_{0} \delta u_{x} EA \epsilon_{0} \: {\rm d}x \\
  \Phi^{(2)}_{0}[\tilde{d}, \delta d]
  & = \int^{L_{0}}_{0}\delta u_{x}
  EA (1 + \epsilon_{0})\tilde{u}_{x} +
 \delta v_{x} EA\epsilon_{0}\tilde{v}_{x} \: {\rm d}x \\
 \Phi^{(3)}_{0}[\hat{d}, \tilde{d}, \delta d]
 & = \int^{L_{0}}_{0} EA\left[
 \delta u_{x}
 (3\tilde{u}_{x}\hat{u}_{x} + \tilde{v}_{x} \hat{v}_{x}) + \delta v_{x} (\tilde{u}_{x} \hat{v}_{x} + 
 \tilde{v}_{x} \hat{u}_{x})
 \right] {\rm d}x \\
 \Phi^{(4)}_{0}[\bar{d}, \hat{d}, \tilde{d}, \delta d]
 & = \int^{L_{0}}_{0}
 EA [\delta u_{x}(3\bar{u}_{x}\hat{u}_{x}\tilde{u}_{x} + \bar{v}_{x}\hat{v}_{x}\tilde{v}_{x} + \bar{u}_{x}\hat{v}_{x}\tilde{v}_{x} + \bar{v}_{x}\hat{u}_{x}\tilde{v}_{x}) + \\
 & \qquad\qquad\quad\:
 \delta v_{x} (3\bar{v}_{x}\hat{v}_{x}\tilde{v}_{x} + \bar{v}_{x}\hat{u}_{x}\tilde{u}_{x} + \bar{u}_{x}\hat{u}_{x}\tilde{v}_{x}+\bar{u}_{x}\hat{v}_{x}\tilde{u}_{x})] {\rm d}x
 \end{aligned}
\end{equation}
where we have set $\Phi^{(n)}_{0} := \left. \Phi^{(n)}(d)\right|_{|d|\rightarrow 0}$, as we refer to the current configuration.

In the rods' \emph{local} frame, the DOFs are collected as $\mathbf{d}_{(l)} = \{u_{a(l)}, v_{a(l)}, u_{b(l)}, v_{b(l)}\}$ (see \autoref{fig:APProdandbeamsketch}(a)), and are transformed to the \emph{global} frame by $\mathbf{d} = (I_{2}\otimes R^{T})\mathbf{d}_{(l)}$, where $R = \left[i_{x}, i_{y}\right]$ is the rotation matrix ($i_{x} = \left\{\cos\theta, \sin\theta\right\}^{T}$, $i_{y} = \left\{-\sin\theta, \cos\theta\right\}^{T}$), and $I_{2}$ is the $2\times 2$ identity matrix.

The rods' natural modes are defined as \cite{argyris-etal_79a_naturalModesFEM}
\begin{equation}
 \label{eq:APProdNaturalModes}
 \mathbf{d}_{\epsilon} = \left\{
 \begin{array}{c} \varepsilon \\ \phi_{r}
 \end{array} \right\} = \left\{
 \begin{array}{c} u_{x} \\ v_{x}
 \end{array} \right\} =
 L^{-1}_{0}\left[
 \begin{array}{cccc}
 -1 &  0 & 1 & 0 \\
  0 & -1 & 0 & 1
 \end{array}
 \right]
 \mathbf{d}_{(l)}
 = N_{0} (I_{2}\otimes R^{T})\mathbf{d}
 = N_{\theta}\mathbf{d}
\end{equation}
and, introducing $\boldsymbol{\chi}^{T}_{1} = \left\{1, 0\right\}$ and $\boldsymbol{\chi}^{T}_{2} = \left\{0, 1\right\}$, such that $u_{x} = \boldsymbol{\chi}^{T}_{1}\mathbf{d}_{\epsilon}$ and $v_{x} = \boldsymbol{\chi}^{T}_{2}\mathbf{d}_{\epsilon}$, the discretization of the first two expressions in \eqref{eq:APPstrainEnergyRod_variations} gives
\begin{equation}
 \label{eq:APPstrainEnergyRod_1&2variations_discretized}
 \begin{aligned}
 \Phi^{(1)}_{0}[\delta\mathbf{d}]
  & = \int^{L_{0}}_{0} (\boldsymbol{\chi}^{T}_{1}N_{\theta}\delta\mathbf{d})^{T}
  EA\epsilon_{0} \: {\rm d}x = \delta\mathbf{d}^{T}N^{T}_{\theta} \boldsymbol{\chi}_{1}\sigma_{0} L_{0} \\
  \Phi^{(2)}_{0}[\tilde{\mathbf{d}}, \delta\mathbf{d}]
  & = \int^{L_{0}}_{0} (\boldsymbol{\chi}^{T}_{1}N_{\theta}\delta\mathbf{d})^{T}EA(1+\epsilon_{0})
  (\boldsymbol{\chi}^{T}_{2}N_{\theta}\tilde{\mathbf{d}}) +
  (\boldsymbol{\chi}^{T}_{2}N_{\theta}\delta\mathbf{d})^{T}EA\epsilon_{0}
  (\boldsymbol{\chi}^{T}_{2}N_{\theta}\tilde{\mathbf{d}}) {\rm d}x \\
  & = \delta\mathbf{d}^{T}N^{T}_{\theta}\left[
  EAL_{0}\boldsymbol{\chi}_{1}\boldsymbol{\chi}^{T}_{1} + 
  \sigma_{0}L_{0}(\boldsymbol{\chi}_{1}\boldsymbol{\chi}^{T}_{1} + \boldsymbol{\chi}_{2}\boldsymbol{\chi}^{T}_{2})
  \right] N_{\theta}\tilde{\mathbf{d}} \\ & = 
  \delta\mathbf{d}^{T} \left[
  N^{T}_{\theta}\mathsf{K_{D}}N_{\theta} + \sigma_{0}
  N^{T}_{\theta}\mathsf{G_{D}}N_{\theta}
 \right]
 \tilde{\mathbf{d}}
 \end{aligned}
\end{equation}
where $\mathsf{K_{D}} := EAL_{0}\boldsymbol{\chi}_{1}\boldsymbol{\chi}^{T}_{1}$, $\mathsf{G_{D}} := L_{0}(\boldsymbol{\chi}_{1}\boldsymbol{\chi}^{T}_{1} + \boldsymbol{\chi}_{2}\boldsymbol{\chi}^{T}_{2})$ are the diagonal stiffness and geometric matrices, referred to the natural coordinates \cite{argyris-etal_79a_naturalModesFEM}. Furthermore, $K := N^{T}_{\theta}\mathsf{K_{D}}N_{\theta}$ and $G := N^{T}_{\theta}\mathsf{G_{D}}N_{\theta}$ give the stiffness and geometric matrices referred to the global coordinates \cite{book:crisfield91}, and we recall that the latter is here independent of the axial stress $\sigma_{0} = EA\epsilon_{0}$, for clarity of the further developments.

With the same formalism, the discretization of the trilinear form in \eqref{eq:APPstrainEnergyRod_variations} reads
\begin{equation}
 \label{eq:APPstrainEnergyRod_3variation_discretized}
 \begin{aligned}
  \Phi^{(3)}_{0}[\hat{\mathbf{d}},\tilde{\mathbf{d}},\delta\mathbf{d}]
  & = \delta\mathbf{d}^{T}N^{T}_{\theta}\left[
  \boldsymbol{\chi}_{1}\left(\tilde{\mathbf{d}}^{T}G\hat{\mathbf{d}} + 
  2\tilde{\mathbf{d}}^{T}N^{T}_{\theta}
  \boldsymbol{\chi}_{1}\boldsymbol{\chi}^{T}_{1}
  N_{\theta}\hat{\mathbf{d}}\right) + 
  \boldsymbol{\chi}_{2}\tilde{\mathbf{d}}^{T}G\hat{\mathbf{d}}\right]EA \\
  \Phi^{(3)}_{0}[\hat{\mathbf{d}},\tilde{\mathbf{d}},\dot{\mathbf{d}}]
  & = \hat{\sigma} \tilde{\mathbf{d}}^{T} G\dot{\mathbf{d}} + 
  \tilde{\sigma} \dot{\mathbf{d}}^{T}G\hat{\mathbf{d}} + 
  \dot{\sigma} \hat{\mathbf{d}}^{T}G\tilde{\mathbf{d}} 
 \end{aligned}
\end{equation}
where the top row, containing the virtual field $\delta\mathbf{d}$ can be used to express the right hand side in \eqref{eq:secondOrderDisplacementCorrection}, whereas the bottom row is used to compute the third order contributions in \eqref{eq:asymptoticCoefficients}.

The quadrilinear form in \eqref{eq:APPstrainEnergyRod_variations} gives a lengthy expression, when keeping all the independent fields. However, we see from \autoref{eq:asymptoticCoefficients} that we only need to compute terms of the form $\Phi^{(4)}_{0}\hat{\mathbf{d}}\tilde{\mathbf{d}}^{3}$, $\Phi^{(4)}_{0}\hat{\mathbf{d}}^{2}\tilde{\mathbf{d}}^{2}$ or $\Phi^{(4)}_{0}\hat{\mathbf{v}}^{4}$, the most general of which can be given the following expression
\begin{equation}
\label{eq:APPstrainEnergyRod_4variation_discretized}
 \begin{aligned}
  \Phi^{(4)}_{0}[\hat{\mathbf{d}}^{2},\tilde{\mathbf{d}}^{2}] = 
  \hat{\sigma}^{2} \tilde{\mathbf{d}}^{T} G\tilde{\mathbf{d}} + 
  4\hat{\sigma}\tilde{\sigma} \tilde{\mathbf{d}}^{T} G\hat{\mathbf{d}} + 
  \tilde{\sigma}^{2} \hat{\mathbf{d}}^{T} G\hat{\mathbf{d}}
 \end{aligned}
\end{equation}

\subsection{Beam element used for frame modeling}
 \label{sApp:beamElement&discretization}

Truss assemblies are modeled according to the above expression, and the strain energy variations \eqref{eq:APPstrainEnergyRod_variations} are accurate at all orders, as the GL strain measure is geometrically exact for a rod \cite{book:antman}.

This is not the case for beams, for which the so-called technical theories, either based on the Green-Lagrange strain, or on the weakly nonlinear approximation (i.e., $\epsilon = \epsilon_{0} + u_{x} + \frac{1}{2}v^{2}_{x}$) introduce spurious strains (see \autoref{sSec:AsymptoticPBanalysisSimpleCases}). Therefore, the approach outlined above, now including the bending modes $\phi_{s}$ and $\phi_{a}$ (see \autoref{fig:APProdandbeamsketch}(b)), associated with the curvature deformation $\chi = v_{xx}$, would lead to an incorrect evaluation of the II order displacement correction and related quantities \cite{byskov_89a_smoothPostbucklingStresses, silvestre-camotim_05a_asymptoticNumericalPostbuckling, NOTE_garcea-etal_13a_postBucklingAsymptotic}.

Then, we need to restore to more advanced beam models, as those outlined, for example, in \cite{garcea-etal_12a_implicitCoRotationalModel, salerno-lanzo_97a_nonlinearFElockingFree, salerno-23a_koitersWorstImperfection, silvestre-camotim_05a_asymptoticNumericalPostbuckling}, to which we refer for further details.

\section*{Acknowledgement}
This research is supported by the Villum Foundation, as part of the Villum Investigator Project ``AMSTRAD'' (VIL54487).

\section*{Conflict of interest}
We have no conflict of interest to declare.

\section*{Replication of results}
The results shown are obtained with an in-house implementation of the FE procedure described in the paper. Further details can be obtained by contacting the authors.


\begin{small}
 \bibliography{biblioMendeley.bib}
\end{small}
\end{document}